\documentclass[12pt,preprint]{aastex}

\def\dfrac{\frac}

\newcommand{\beq}{\begin{equation}}
\newcommand{\eeq}{\end{equation}}
\newcommand{\bea}{\begin{array}}
\newcommand{\eea}{\end{array}}

\shorttitle{Hot Jupiter} \shortauthors{Zhou et
al.}

\begin{document}

\title{Interaction of Close-in Planets with the Magnetosphere of their
Host Stars I: Diffusion, Ohmic Dissipation of Time Dependent Field, 
Planetary Inflation, and Mass Loss}

\author{Randy O. Laine\altaffilmark{1},  
Douglas N.C. Lin\altaffilmark{2,3}, \&
Shawfeng Dong\altaffilmark{2}}
\altaffiltext{1}{Ecole Normale Superieure, Paris, France, randy.laine@ens.fr}
\altaffiltext{2}{UCO/Lick Observatory, University of California, Santa Cruz, CA 95064, USA lin@ucolick.org, dong@ucolick.org}
\altaffiltext{3}{Kavli Institute of Astronomy \& Astrophysics, Peking University, Beijing,  China}

\begin{abstract} 
The unanticipated discovery of the first close-in planet around 51 Peg
has rekindled the notion that shortly after their formation outside
the snow line, some planets may have migrated to the proximity of
their host stars because of their tidal interaction with
their nascent disks. After a decade of discoveries, nearly 20 \% of
the 200 known planets have similar short periods. If these 
planets indeed migrated to their present-day location, their survival
would require a halting mechanism in the proximity of their host
stars. Here we consider the possibility that a magnetic coupling
between young stars and planets could quench the planet's orbital
evolution. Most T Tauri stars have magnetic fields of several thousand
gausses on their surface which can clear out a cavity in the
innermost regions of their circumstellar disks and impose
magnetic induction on the nearby young planets. After a brief
discussion of the complexity of the full problem, we focus our
discussion on evaluating the permeation and ohmic
dissipation of the time dependent component of the stellar magnetic
field in the planet's interior. Adopting a model first introduced by
Campbell for interacting binary stars, we determine the
modulation of the planetary response to the tilted magnetic field of a
non-synchronously spinning star. We first compute
the conductivity in the young planets, which indicates
that the stellar field can penetrate well into the planet's envelope in
a synodic period.  For various orbital configurations, we show that the
energy dissipation rate inside the planet is sufficient to induce
short-period planets to inflate. This process
results in mass loss via Roche lobe overflow and in the halting of the
planet's orbital migration.
\end{abstract}

.

\section{Introduction}
Perhaps the most surprising finding in the search for extrasolar
planets is the discovery of short-period ($P < 1$ week) Jupiter-mass
($M_J$) companions around Solar-type main sequence stars (Mayor \&
Queloz 1995, Marcy et al. 2000).  Among the inventory of $>$200
presently-known extrasolar planets, 20\% have $P = 1-7$ days. Nearly
20 short-period planets have measured radii ($R_p$) that are comparable to or
larger than that of Jupiter ($R_J$). While these information may be
biazed because of observational selection effects, these planets are
most probably gas giants.

According to the conventional sequential-accretion scenario (Pollack
{\it et al.} 1996), the most likely birth place for gas giant planets
is just outside the snow line where volatile heavy elements can
condense and coagulate into large planet building blocks (Ida \& Lin
2004). In protostellar disks with comparable surface density
($\Sigma$), metallicity ([Fe/H]), and temperature ($T$) distributions
as those of minimum mass nebula model (Hayashi {\it et al.} 1985),
protoplanets with $M_p \sim M_J$ induce the formation of a gap near
their orbit as a consequence of their tidal torque on the nascent
disks (Goldreich \& Tremaine 1978, 1980, Lin \& Papaloizou 1980,
1986a, 1993).  In relatively massive and fast evolving disks, the
outward transfer of angular momentum due to the disks' intrinsic
turbulence can lead to an inward mass flux ($\dot M_d$) and the
migration of the gas giant planets (Lin \& Papaloizou 1986b). This
process is commonly referred to as type II migration (Ward 1997).

This migration scenario was resurrected to account for the origin of
the first known short-period extra solar planet (Lin {\it et al.}
1996).  Although type-II migration provides a natural avenue for
relocating some gas giants, a mechanism is needed to retain these
planets close to their host stars. Moreover, many stars are born with
rapid rotation (Stassun 2001). When young planets venture close to
their host stars, angular momentum would be transferred from the
stellar spin to the planet's orbit if the stellar spin frequency
$\omega_\ast$ is still larger than the planet's orbital frequency
($\Omega_k$).  The rate of the star-to-planet angular momentum
transfer intensifies rapidly and may exceed that from the planet to
the disk.

Two basic physical effects were suggested as potential migration barriers.
The first one is tidal interaction between the host star and the planet. The
gravitational perturbation of the star and close-in planet leads to
responses in both the star and planet. For tidal frequencies smaller than twice the
spin frequency, inertial waves are excited in the convective envelope
of the host star and are dissipated there by turbulent viscosity
(Ogilvie \& Lin 2007).  But, the tide excited by a close-in gas giant
planet in a star, with a structure similar to that of the present Sun,
marginally fails to achieve nonlinearity so that their survival is
ensured. Nevertheless, during the formation epoch of solar-type stars,
conditions at the center of the star evolve, so that nonlinearity may
set in at a critical age, resulting in a relatively intense
star-planet tidal interaction.

The second effect suggested is based on the magnetic interaction 
between the host star and the planet. 
Young T Tauri stars also have radii ($R_\ast$) 2-3 times that of the
present-day Sun ($R_\odot$) and several thousand gausses fields
($B_\ast$) on their surface (Johns-Krull 2007). The stellar magnetosphere
threads across the inner regions of the disk and clears a cavity out
to a critical radius ($R_c$) which is determined by both the magnitude
$B_\ast$ and $\dot M_d$ (Konigl 1991). The subsequent complex interplay
between accretion and outflow leads to angular momentum exchange which
induces $\omega_\ast$ to evolve toward $\Omega_k$ at $R_c$ (Shu 1994).
When the planet's orbital semi major axis ($a$) reduces well inside
$R_c$, its Lindblad resonances relocate inside the star's
magnetospheric cavity. In principle, the planet's migration would
stall due to its diminishing tidal torque on the disk.

However, if the star's magnetospheric interaction with the disk can
lead to $\omega_\ast = \Omega_k (R_c)$, the planet inside the
magnetospheric cavity would have $\Omega_k > \omega_\ast$.  In this
limit, the star-planet tidal interaction would induce a transfer of
angular momentum from the planet to the star. In addition, the
differential motion between the planet and the stellar spinning
magnetosphere induces an electromagnetic field with a potential to
generate a large current analogeous to the interaction between the Jovian
magnetosphere with its satellite Io (Goldreich \& Lynden-Bell 1969).
The associated Lorentz force drives an orbital evolution toward a
synchroneous state, in which case, angular momentum would be
transferred from the planets with $\Omega_k > \omega_\ast$ to their
host stars, and the planets would continue their orbital decay.

In order to determine the necessary condition for the retention of
close-in young planets, we examine, in this paper, their interaction
with the magnetosphere of their host T Tauri stars.  
In \S2, we briefly recapitulate the essential concepts and validity of previous
investigations on some related topics and give an overview of the key 
phenomena that will be discussed in this and later papers. In \S3, we
adopt an existing model in order to examine the interaction between a
planet and the magnetosphere of its host star. In \S4, we compute
precisely the planet's magnetic diffusivity for a specific set of
parameters, as well as the corresponding ohmic dissipation rate within
that planet. In \S5, we suggest that the ohmic dissipation can
generate sufficient heat to inflat the planet.  In \S6, we construct an idealized 
self-consistent model in which the polytropic and isothermal equations
of state are utilized.  These equations represent the expected outcome
of radiation transfer within the fully convective interior and the
isothermal surface of a close-in planet which is exposed to the
intense radiation from its host star.  This internal model allows us
to compute the magnetic diffusivity.  With these tools, we discuss in
\S7 the structural adjustment of the planet in response to this
heating source and we compute the ohmic dissipation and mass loss rate
for different set of parameters. Finally in \S8, we summarize our
results and discuss their implications.

\section{Planetary and astrophysical analogue}

Two previous analyses are directly relevant to the present study: 1) the
interaction of Io with the magnetosphere of Jupiter and 2) the spin-orbit
synchronization in binary stars containing a magnetized white dwarf
and its main sequence or white dwarf or planetary companion. 

\subsection{Unipolar Induction in Io}
Io orbits around Jupiter inside its magnetosphere once every 1.7 days
which is considerably longer than Jupiter's 10 hours spin period. This
relative motion imposes a periodic variation in Jupiter's decametric
emission (Duncan 1966).  A class of models that accounts for the
origin of this emission was developped based on the assumption that Io
has a sufficiently high conductivity. In Io's rest frame, the
electric field vanishes and the steady component of Jupiter's magnetic
field permeates in Io's interior over time. When a steady state is established, 
the tube of constant magnetic flux is firmly frozen into Io 
(Piddington and Drake 1968) due to its high conductivity. 
The flux tube carried by Io moves through the surrounding field lines (which
corotate with Jupiter) and slips through Jupiter's less conductive ionosopheric
surface. Plasma in Jupiter's ionosphere flows around the tube and
introduces a potential difference across it (Goldreich \& Lynden-Bell
1969). The associated electric field drives a current which 
travels down one half of the flux tube from Io and is sent back to Io
along the other half. Within the flux tube connecting Io and Jupiter's
ionosphere and those across it on Io, the electric field vanishes as a
consequence of high conductivity. Thus, this DC circuit is closed by
Io as a unipolar inductor.

The magnitude of the electric current is primarily determined by the Pederson
conductivity at the foot of the flux tube {\it i.e.} on the ionosphere
of Jupiter. Finite conductivity also determines the magnitude of the
drag against the slipage of the flux tube through Jupiter.  This drag
results in energy dissipation on the surface of Jupiter and in a
torque on the orbital motion of Io, driving the system towards a state
of synchronization.  This configuration is justified by the assumption
that a constant flux tube is firmly anchored on and dragged along by
Io which requires the conductivity in Io to be much higher than that
in Jupiter's ionosphere. A 10$^o$ inclination between Jupiter's
magnetic dipole and rotation axis does introduce a periodic variation
(over a synodic period) in the field felt by Io. The permeation and
dissipation of this time-dependent AC field may be negligible in the
limit of high conductivity in Io.

The validity of the key assumption for the unipolar induction model
({\it i.e.} conductivity on Io is larger than that in Jupiter's
ionosphere) has also been challenged by Dermott (1970).  A modest
resistence in Io would distort the field which may lead to field
slippage through Io.  In this case, the passage of Io through the
magnetosphere of Jupiter would lead to the generation of Alfven waves
along the flux tube (Drell {\it et al.}  1965, Neubauer 1980).  But,
due to the field displacement, the waves, partially reflected at the
foot of the flux tube on Jupiter's surface, may not be able to return
to Io, in which case the DC circuit would be broken and the motion of
Io would be decoupled from the that of the flux tube.  Nevertheless,
the Alfven waves are dissipated inside both Io and Jupiter, leading to
a torque which must depend on their penetration depth.

An alternative class of scenarios has been proposed based on the
assumption that the magnetosphere is everywhere anchored on Jupiter
and the flux tube moves freely through Io (Gurnett 1972).  This model
requires the conductivity in Jupiter's ionosphere to be larger than
that in Io.  It assumes that the presence of Io creates a plasma
sheath with an electric field to cancel the induced EMF associated
with the motion of Io relative to Jupiter's magnetosphere (Shawhan
1976).  The simplifying approximations in the development of this
theory have been challenged by Piddington (1977) who questioned both
the validity of the sheath-creation mechanism and the self-consistency
of the internal and external field configurations, subjected to the
electric currents in and around Io.

On the observational side, UV emissions from Io's footprint on
Jupiter has been observed. But, it extends
well beyond the intersection between Io's flux tube and Jupiter's
ionosphere and the emssion downstream is protracted (Clarke {\it et
al.} 1996). These observations do not agree with the simple
interpretation of either the unipolar induction or the plasma sheath
scenarios.

\subsection{Magnetic Coupling in Interacting Binary Stars}
There are many close binary-star systems with a white dwarf as their
primary component.  These systems also contain main sequence stars and
other white dwarfs as secondary components in compact and circular
orbits around each other.  In some cases, mass is transferred from the
secondary to the primary.  In other cases, gravitational radiation may
play an important role in determining the evolution of these systems.

A sub class of such interacting binary stars, AM Her systems, is
composed of a magnetized white-dwarf primary and a lower-mass main
sequence star as its secondary in a fully synchroneous orbit despite
the ongoing mass transfer between them (Warner 1995).  This orbital
configuration is very similar to that of the Jupiter-Io system despite
the enormous difference between the mass ratio in the two cases. The
motivation for studying the impact of magnetic coupling between these
stellar components is to assess whether this synchronous state can be
achieved through the ohmic dissipation of the white dwarf's field in
the main sequence star's surface (Joss {\it et al.} 1979).  Toward
this goal, Campbell (1983, hereafter C83) adopted a novel approach by
considering the penetration and dissipation of a periodically variable
field, associated with an asynchronously spinning primary. 

Campbell's approach is fundamentally different from that of the
unipolar induction model.  In this analysis, Campbell focussed on the
flow in the envelope of the secondary and neglected the possibility of
current flowing through the flux tube between the secondary star/satellite
and the surface of the primary star/planet. This vacuum-surrounding
approximation is justifiable since conductivity on the primary is
likely to be much larger than that on the secondary and 
the stationary component of the field is
frozen in the white-dwarf primary but not in the main-sequence
secondary. Campbell analyzed the time dependent response of
the secondary, including the modification of the field by the induced
(AC) current in it (Campbell 2005), to the periodic modulation of the
field.  In contrast, the unipolar induction model depends on the
explicit assumption that the field is anchored on the secondary and
its distortion near the secondary must be small so that a complete
current loop can be established between the primary and the
secondary. Campbell determined the periodic diffusion of the field and
the ohmic dissipation of the induced AC current in the companion
whereas that of the induced DC current is assumed to occur in the
primary in the unipolar induction model.

In recent applications of the unipolar induction model in the context
of interaction between white dwarf binary stars, the current induced
by the unperturbed field has been computed but the induced field
generated by the current was neglected (Wu {\it et al.}  2002,
Dall'Osso {\it et al.}  2006). A tidal torque is computed at the
footprint of the flux tube, which is attached to the secondary white
dwarf, on the surface of the magnetized primary white dwarf.  A
totally self-consistent solution of this diffult and complex problem
remains outstanding.  In addition to the uncertain anchorage location
of the field, it is not clear whether the resulting misalignment of
the total (original plus induced) field and current may be
sufficiently large to break the circuit, in which case, Campbell's
model may be more appropriate.

\subsection{Mathematical approximations made by the two previous models}
In this subsection, we summarize the physical description of the two
models just presented (the unipolar inductor versus the periodic
diffusion) by explicating the mathematical approximations made in each
one of them.  The complete MHD induction equation can be expressed as
\begin{equation}
\frac{\partial \textbf{B}}{\partial t}=
\nabla \wedge (\upsilon\wedge 
\textbf{B})
-\nabla \wedge (\eta \nabla \wedge\textbf{B})
\label{complete mhd induction equation}
\end{equation} 
where the magnetic diffusivity $\eta=1/\mu_{0}\sigma$ and the
electrical conductivity $\sigma$ functions of position and $\mu_{0}$
is the permeability.  If $\eta$ is constant, the second term on the
right hand side becomes $-\nabla \wedge (\eta \nabla
\wedge\textbf{B})= \eta\ \nabla^{2} \textbf{B}$ which reduces to a
common expression of diffusion.  The two models (unipolar induction
versus periodic diffusion) consider two complementary approximations
of equation \ref{complete mhd induction equation}.  In the problem
where Io is treated as an unipolar inductor, its conductivity is
explicitly assumed to be large so that the second term on the right
hand (the diffusion term) is negligible compared to the first ({\it
i.e.} the induction term).  In this configuration, one can show that
the field lines of the steady component of the magnetic field are
moving with Io and appear to be "frozen" on Io (see appendix A).
Alternatively, in the model considered by Campbell, it is the first
term on the right hand side that is being neglected. Moreover, only
the diffusion of the time dependent component of the field is being
considered.  This approximation is valid if the two interacting bodies
are almost in corotation ({\it i.e.} the relative speed $\upsilon$
that appears in equation (\ref{complete mhd induction equation}) is
small), or if the conductivity in the secondary is small.

\subsection{Overview of the phenomena that will be discussed}
\label{subsection overview of our model}
The process under investigation in this paper is analogeous to both
the Jupiter-Io and the interacting binary star problems.  In fact, the
unipolar induction model has already been applied to study the orbital
evolution of terrestrial planets or the cores of gas giants around
white dwarfs (Li {\it et al.} 1998). There are even follow-up
determination of the radio flux densities from potential
white-dwarf/planet systems (Willes \& Wu 2005). In this analysis,
although the dissipation of the induced current due to the finite
conductivities in the white dwarf was considered, the feedback
modification of the field and the dissipation within the planet have
been neglected (Li {\it et al.} 1998).  As discussed above, it is not
clear whether a DC circuit can be closed to promote the unipolar
induction mechanism.

In light of these uncertainties, we consider both classes of models
for the interaction of close-in planets with their magnetized host
stars. In this paper, we focus our discussion on the
mechanism described by Campbell, 
and apply it to a hot-jupiter revolving around its star. We
will return to the unipolar induction problem in a later paper.

When young planets first arrive at the vicinity of their host stars,
they are unlikely to be in a totally synchronized state.  The stellar
magnetic field felt by the planet may be dominated by the periodic
modulation associated with the synodic (between the stellar spin and
the planet's orbit) motion. In addition, the temperature in the
planet's surface is expected to be $\sim 10^3$ K and the conductivity
there may be moderate.  In response to the modulation of the field,
the interior of the planet continually adjusts to the magnetization
effects so that the flux tube cannot be effectively frozen in the
planet.  All of these boundary condition suggest that at least over
some regions of the planet (especially on the night side where the
photo ionization due to the stellar flux is negligible), the
modulation of the field may lead to an induced current inside the
planet which does not contribute to the close circuit of a unipolar
inductor.

Following the geometry introduced by Campbell (C83), we consider a
close-in gas giant planet, with a finite conductivity, interacting
with a time-dependent magnetic field generated by the star. An induced
current is generated inside the planet, which is associated with an
ohmic dissipation rate. Our main contributions to the model used by
Campbell are: 1) the relevant diffusivity inside the gas giant
planets, 2) the effects of the ohmic dissipation on the planet's
internal structure, and 3) the resulting orbital evolution of the
planet. (Items 2 and 3 have negligible consequence in the interacting
binary star problem considered by Campbell). Since we are only
considering the dissipation in the planet's interior, the associated
torque applied on its orbit should be regarded as a lower limit.

In our scenario, we postulate that at sufficiently close proximity to
the host star, the stellar magnetic field is sufficiently intense that
the ohmic dissipation of the periodically diffused field inside the
planet is adequate to heat and to inflat the planet until it overflows
its Roche lobe. The hemisphere of the planet facing its host star is also exposed to
the intense flux of UV radiation during the stellar infancy. It is
possible for the planet to develop a substantial ionosphere regardless
the state of synchronization between the planet's orbit and spin (the
time scale for establishing local ionization equilibrium is much
faster than the planet's spin and orbital periods).  

We will separately study these two phenomena (angular momentum transfert due 
to mass loss and presence of a ionosphere) in the follow-up papers of this serie. 
We will show that the angular momentum transfer associated with the mass transfer can halt the
orbital evolution of the planet. We will also present an analysis on the conductivities in the planet's day-side
ionosphere and on the host stars surface. This will lead to an
analysis on the condition for the unipolar induction to effectively
operate and apply a significant slow-down torque on the planet's
orbit.

\section{Magnetic induction} 
\label{chapter magnetic induction}

In this section, we are going to derive the governing equations that
we use to compute the ohmic dissipation rate. Various equations are
presented here for the purpose of introducing the algorithm of the
numerical models to be presented in subsequent chapters.  Although we
follow closely the approach made in C83, for brevity, we do not
repeatedly cite this reference. But, wherever similarities occur,
referal of Campbell's earlier work is implicitly implied. Also, throughout the paper, we use SI units. 

\begin{figure}[htbp]
\begin{center}
\epsscale{0.5}
\plotone{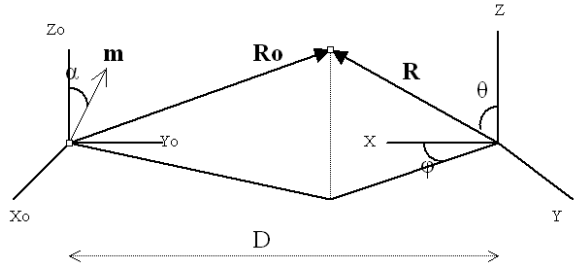}  
\end{center}
\caption{The geometry of the system. The star is on the left, at the center of the 
set of axies ($x_{0}$,$y_{0}$,$z_{0}$), and the planet is on the right, at the center of 
the set of axies (x,y,z).}
\end{figure}

We consider a protoplanetary system with a gas giant planet revolving
around a T Tauri star with an angular frequency $\Omega_p$. Well
beyond the planet's semi major axis, there is also a protoplanetary
disk. The host star has a dipolar moment $\textbf{m}$
tilted with an angle $\alpha$ with respect to its spinning axis (see figure 1). 
The angular frequency of the stellar spin is $\omega_\ast$.  The orbital
axis of the disk and the planet are parallel to the star's spinning
axis. The following analysis is applied to a frame of reference
centered on the star and rotating with the planet. 

In this frame, the planet is a fixed object (the planet's spin is
neglected) in a periodic magnetic field with a frequency $\omega =
\omega_\ast -\Omega_p$.  From Ohm's law 
$\textbf{J}=\sigma \textbf{E}$ 
and Maxwell's equations $\dfrac{\partial \textbf{B}}{\partial t} = 
-\nabla \wedge \textbf{E}$ and 
$\nabla \wedge \textbf{B} = 
\mu_{0} \textbf{J}$, 
the equation on the magnetic field becomes:
\begin{equation}
\dfrac{\partial\textbf{B}}{\partial t}
=- \nabla\wedge(\eta \nabla \wedge\textbf{B}) .
\label{B time dependent}
\end{equation}

It follows that in the mechanism considered by Campbell 
(as well as in this paper), it is the time dependent stellar 
magnetic field, diffusing inside the 
secondary (for Campbell) or the hot Jupiter (in our paper), 
as well as the planet's induced magnetic field, 
that generate the current inside the planet, following the equation 
$\nabla \wedge \textbf{B} = 
\mu_{0}\textbf{J}$.
The relative speed between the planet and the stellar magnetic 
field thus intervenes not through 
$\textbf{E} = 
-\upsilon \wedge \textbf{B}$ 
but through the time dependence in the stellar magnetic field that diffuses in the planet. 

Following C83, we only consider the poloidal component $\phi$ of the magnetic field: 
\begin{equation}
\textbf{B}\ =\  
\nabla\wedge(\nabla\wedge(\phi\ \mathbf{e_{r}}))
\label{definition poloidal scalar}\\
\end{equation}
where $\phi$ is a function of r, $\theta$, $\varphi$. and can be expanded 
in terms of the sperical harmonics
$Y_{l}^{m}(\theta,\varphi)$ (equation \ref{decomposition of
Phi}). Moreover the variation in time of the magnetic field felt by
the planet is periodic. In the limit where the field penetrates
quickly in the planet compared to the time scale on which the field
changes (so that the planet can respond ``adiabatically''), we can
account for the time dependence of $\phi$ by mutiplying its spatial
part by $e^{i\omega t}$:
\begin{eqnarray}
\phi(\textbf{r},t)
=\mu_{0}\left[\sum_{l,m}\ C_{l}^{m}\
G_{l}(r)\ Y_{l}^{m}(\theta,\varphi)\right]e^{i\omega t}
\label{decomposition of Phi} \ \ \ \ \ \ \ (l \geqslant 0\ and\
-l\leqslant m\leqslant l)
\end{eqnarray}
where $C_{l}^{m}$ are constant coefficients and $G_{l}(r)$ is a
function of $r$ to be determined. We then replace \textbf{B} in 
the left hand of (2) by its expression in (3). After integration, we obtain 
\begin{equation}
\nabla\wedge\mathbf{B}\
=-\dfrac{i\omega}{\eta}\nabla\wedge
\left(\phi\ \mathbf{e_{r}}\right).
\label{rhs=lhs}
\end{equation}
We then replace \textbf{B} in the left hand side of this equation using (\ref{definition poloidal scalar}), and 
develop both sides of the equation. After identification, we obtain: 
\begin{equation}
\dfrac{d^{2}G_{l}}{dr^{2}}(r)-\left(\dfrac{l(l+1)}{r^{2}}+
\dfrac{i\omega}{\eta}\right)G_{l}(r)=0 \ \ \ \ \ \ \ \ \ \ \ 
{\rm inside} \ {\rm the} \ {\rm planet.} \label{first equation on G}
\end{equation}
This equation holds inside and outside the planet (same as eqs 16 and
18 in C83). However, outside the planet, the conductivity is assumed
to be very low and, therefore, the magnetic diffusivity is extremely
high compared to the diffusivity inside the planet. In the limit where
the magnetic diffusivity outside tends to infinity (equivalent to a
vacuum surrounding), the equation (\ref{first equation on G}) becomes:
\begin{equation}
\dfrac{d^{2}G_{l}}{dr^{2}}(r)-\left(\dfrac{l(l+1)}{r^{2}}\right)G_{l}(r)
=0 \ \ \ \ \ \ \ \ \ 
\ \ \ \ \ \ \ \ \ \ {\rm outside} \ {\rm the} \ {\rm planet.} \label{second equation on G}
\end{equation}

We consider the radial part of the poloidal scalar outside the planet. Following C83,  
we introduce $\phi_{star}$ the radial part of the poloidal scalar outside the planet 
due to the star's magnetic field, and $\phi_{planet}$ the radial part of the poloidal scalar 
outside the planet due to the planet (cf C83 eqs 21-22): 
\begin{eqnarray}
\phi_{star}=\dfrac{\mu_{0}m {\rm \sin}\alpha}{8\pi d^{3}}r^{2} 
\left(2{\rm cos}\varphi\ {\rm \sin}\omega t+\ {\rm \sin}\varphi\ 
{\rm cos}\omega t\right)P_{1}^{1} 
+ \dfrac{\mu_{0}m {\rm \sin}\alpha}{8\pi\ d^{4}}r^{3}\left[P_{2}^{0}
{\rm \sin}\omega t \right.   \nonumber \\
-\left. \left(\dfrac{1}{2} {\rm cos}2\varphi\ {\rm \sin}\omega t\ 
+ \dfrac{1}{3}{\rm \sin} 2\varphi\ {\rm cos}\omega t\right)P_{2}^{2}\right] 
\label{phi_star outside}
\end{eqnarray}

\begin{eqnarray}
\phi_{planet}=\mu_{0}P_{1}^{1}\left[\dfrac{{\rm cos}\varphi}{r} 
(\alpha_{1}\ {\rm \sin}\omega t\ +\alpha_{2}\ {\rm cos}\omega t)
+\dfrac{ {\rm \sin}\varphi} {r}(\alpha_{3}\ {\rm \sin}\omega t\ 
+\alpha_{4}\ {\rm cos}\omega t)\right]\ 
\nonumber \\
+\ \dfrac{\mu_{0}P_{2}^{0}}{r^{2}}(\beta_{1}\ {\rm \sin}
\omega t\ +\ \beta_{2}\ {\rm cos}\omega t) 
+\mu_{0}P_{2}^{2}\left[\dfrac{ {\rm cos}2\varphi}
{r^{2}}(\gamma_{1}\ {\rm \sin}\omega
t\ +\ \gamma_{2}\ {\rm cos}\omega t)
 +\dfrac{{\rm \sin}2\varphi}{r^{2}}(\gamma_{3} {\rm \sin}\omega t\ 
+\ \gamma_{4}\ {\rm cos}\omega t)\right]
\label{phi_planet outside}
\end{eqnarray}
where $P_{1}^{1}=- {\rm \sin}\theta,\ P_{2}^{0}=\frac{1}{2}(3 {\rm cos}^{2}
\theta-1)$, and $P_{2}^{2}=3 {\rm \sin}^{2}\theta$ are the associated 
Legendre polynomials (our convention for $P_{1}^{1}$ has an opposite 
sign as that adopted by Campbell). In addition, $\phi_{planet}$ has the 
same time and angular dependence as $\phi_{star}$ because the field 
inside the planet is induced by the stellar's magnetic field. 

The sum $\phi_{outside}=\phi_{star}+\phi_{planet}$ is the total
poloidal scalar outside the planet, and
$\phi_{outside}$ (given by (\ref{phi_star outside}) and
(\ref{phi_planet outside})) is equal to $\phi_{inside}$ (given by
(\ref{decomposition of Phi})) at the surface of the planet
($r=R_{p}$).

\subsection{Poloidal scalar inside the planet}
\label{subsection poloidal scalar inside}
In order to determine the poloidal scalar inside the planet, we first
numerically calculate the values of $G_{l}(r)$ (the radial part of
$\phi$, cf. equation \ref{decomposition of Phi}) and $G_{l}'(r)$
inside the planet by solving equation (\ref{first equation on G}). We
then calculate the coefficients $C_{l}^{m}$, which appears in the
decomposition of~$\phi$. They are determined by the boundary
conditions which connect the interior and exterior solutions.  In the
rest of this section (\S3), we assume that the conductivity profile is
known, and we describe the procedure used to compute the ohmic
dissipation rate inside the planet.  In the following sections, we
apply the method described in \S3 to compute the ohmic dissipation
rate inside the planet.

In the following sections, we compute the magnetic dissipation and
numerically obtain the value of the ohmic dissipation rate.

\subsubsection{Computation of G(r)}
If the diffusivity $\eta(r)$ is known, we can solve (\ref{first
equation on G}) numerically, for $l=1$ and $l=2$, with a two-point
boundary solver using the Newton-Raphson-Kantorovich method, and the
equations and boundary conditions are given below:
\begin{eqnarray}
\left\{
\begin{array}{l}
Y_{1}(r)=\mathcal{R}e(G(r))\\
Y_{2}(r)=\mathcal{I}m(G(r))
\end{array}
\right.
\left\{
\begin{array}{l}
{\rm Equations:} \\
Y'_{1}(r)=Y_{3}(r) \\
Y'_{2}(r)=Y_{4}(r) \\
Y'_{3}(r)=-\dfrac{\omega}{\eta(r)}Y_{2}(r)+\dfrac{l(l+1)}{r^{2}}Y_{1}(r) \\
Y'_{4}(r)=\dfrac{\omega}{\eta(r)}Y_{1}(r)+\dfrac{l(l+1)}{r^{2}}Y_{2}(r)
\end{array}
\right.
\left\{
\begin{array}{l}
{\rm Boundary} \ {\rm conditions} \ {\rm at} \ r=R_{p}\ \& \ 0 \\
G'_{l}(R_{p})+\dfrac{l}{R_{p}}G_{l}(R_{p})-(2l+1)R_{p}^{l}=0\\
G'_{l}(r\simeq 0)-\dfrac{l+1}{r}G_{l}(r\simeq 0)=0 \label{detailed equations}
\end{array}
\right.
\end{eqnarray}

\subsubsection{Computation of the C(l,m)}
The complex coefficients $C_{l}^{m}=\mu_{l}^{m}+i\nu_{l}^{m}$ have
real and imaginary parts $\mu_{l}^{m}=Re(C_{l}^{m})$ and
$\nu_{l}^{m}=Im(C_{l}^{m})$. We equate the real part of the
decomposition of the poloidal scalar inside the planet given in
(\ref{definition poloidal scalar}) at $r=R_{p}$ (radius of the planet)
with the expression of $\phi_{outside}=\phi_{star}+\phi_{planet}$
given in (\ref{phi_star outside}) and (\ref{phi_planet outside}) at
$r=R_{p}$.

Moreover, using the fact that ($P_{1}^{1}$,$P_{2}^{0},P_{2}^{2}$) and
then (${\rm cos} \omega t {\rm cos} \varphi$, ${\rm cos} \omega t {\rm \sin}
\varphi$, ${\rm \sin} \omega t {\rm cos}\varphi$, ${\rm \sin}\omega t
{\rm \sin}\varphi$) (${\rm cos} \omega t$,${\rm \sin}\omega t$),
(${\rm cos}\omega t {\rm cos} 2\varphi$, ${\rm cos} \omega t {\rm \sin}
2\varphi$,${\rm \sin} \omega t {\rm cos} 2\varphi$, ${\rm \sin} \omega t
{\rm \sin} 2\varphi$) are a set of bases, we get a
set of linear equations which can be solved for
($\mu_{1}^{1}$, $\mu_{1}^{-1}$, $\nu_{1}^{1}$, $\nu_{1}^{-1}$,
$\alpha_{1}$, $\alpha_{2}$,$\alpha_{3}$,$\alpha_{4}$), ($\mu_{2}^{0}$,
$\nu_{2}^{0}$, $\beta_{1}$,$\beta_{2}$), and ($\mu_{2}^{2}$,
$\mu_{2}^{-2}$, $\nu_{2}^{2}$, $\nu_{2}^{-2}$, $\gamma_{1}$,
$\gamma_{2}$, $\gamma_{3}$, $\gamma_{4}$) (the linear systems verified
by these unknowns are given in appendix B).

\subsection{Computation of the ohmic energy dissipation rate}
\label{computation of the ohmic dissipation rate}
The potential generates an electric field $\mathbf{E}$ which induces a
volumic current $\mathbf{J}$ inside the planet.  The associated ohmic
dissipation inside the planet is
$\mathcal{P}_{volumic}=Re(\mathbf{J})\ Re(\textbf{E})$.  Using
$\textbf{E}=\dfrac{1}{\sigma} \mathbf{J}$ and
$\mathbf{J}=\frac{1}{\mu_{0}} \nabla\wedge\textbf{B}$, we can write:
\begin{equation}
\mathcal{P}=\int_V \dfrac{1}{\sigma(r)} \left(\mathcal{R}e(\mathbf{J})
\right)^{2}dV=\int_V \dfrac{1}{\sigma(r)} \left[\mathcal{R}e
\left(\dfrac{\nabla\wedge\textbf{B}}{\mu_{0}}\right)\right]^{2}.  dV
\label{W provisoire}
\end{equation}
Moreover, using equation (\ref{rhs=lhs}), we can write
\begin{eqnarray}
\mathcal{P}=\dfrac{\omega^{2}}{\mu_{0}}\int
\dfrac{1}{\eta(r)}\left[\dfrac{1}{ {\rm \sin} \theta}\left(\dfrac{\partial
Im(\phi)}{\partial \varphi}(r, \theta,\varphi)\right)^{2} + {\rm \sin}
\theta\left(\dfrac{\partial Im(\phi)}{\partial \theta}(r, 
\theta\varphi)\right)^{2}\right]dr\ d\theta\
d\varphi. 
\label{dependence of P}
\end{eqnarray}
We use eq. (\ref{decomposition of Phi}) to express the real and
imaginary parts of $\Phi$.  After integrating over $\theta$ and
$\varphi$, we are left with:
\begin{equation}
\mathcal{P}= \int_r 
< \mathcal{P}_{volumic} > r^2 dr
\label{energy_integrand_expression}
\end{equation}
where the angle-integrated volumic power and  
\begin{footnotesize}
\begin{eqnarray}
< \mathcal{P}_{volumic} > =
\dfrac{\mu_{0} \omega^{2}}{\eta r^2 }\left\{ {\rm cos}^{2}\omega t
\left[\left(A_{11}^{2}+A_{12}^{2}\right)
+3\left(A_{17}^{2}+A_{18}^{2}\right)
+\dfrac{3}{\pi} A_{15}^{2}\right]
\right. \nonumber \\ \left.
+{\rm \sin}^{2}\omega t\left[\left(A_{13}^{2}
+A_{14}^{2}\right)+3\left(A_{19}^{2}
+A_{20}^{2}\right)+\dfrac{3}{\pi}A_{16}^{2}\right] \right. 
\nonumber \\ \left.
+ {\sin \omega t} {\cos \omega t}
\left[\left(A_{12}A_{14}+
A_{11}A_{13}\right)+3
\left(A_{17}A_{19}+A_{18}A_{20}\right)+\dfrac{3}{\pi}
A_{15}A_{16}\right]\right\} \nonumber \label{energy integral}
\end{eqnarray}
\end{footnotesize}
where the expressions for $A_{ij}$ are given in Appendix C. 

\section{Conductivity profile and ohmic dissipation rate}
\label{shawfeng's conductivity}
The general setting of the problem and the basic equations have been
laid down. We have seen that once a conductivity profile is chosen,
one can solve (\ref{detailed equations}) and determine $G_{l}(r)$
inside the planet (the radial part of the poloidal scalar inside the
planet). Then, one can compute the $C(l,m)$, and finally obtain the
ohmic dissipation rate $\mathcal{P}$ inside the planet.

\subsection{Computation of $\mathcal{P}$ for one specific set of parameters} 
We compute the conductivity inside the planet with two parallel
approaches. In \S\S 5-7, we develop an idealized self-consistent internal
structure model to determine the response of the planet to the ohmic
dissipation of the induced current in it. But, in this section, we
first introduce a realistic, but non self consistent, model with the
following set of parameters:
\begin{flushleft}
\textbf{Planet's mass and radius:} $0.63 M_J$ 
\& $R_{p}=1.4\ R_{J}=10^{8}\ m$, \\
\textbf{Semi-major axis:} $a=0.04\ AU=6\times 10^{9}\ m$. \\
\end{flushleft}
These and other (such as mass and luminosity) stellar parameters are
appropriate for the short-period planet around HD209458 (Bodenheimer
{\it et al.} 2001).  We compute the internal conductivity due to the
ionization of the alkaline metals (see Appendix D for details).
Although the planet is heated on the day side, thermal circulation can
redistribute the heat and reduce the temperature gradient between
the two side of the planet (Burkert {\it et al.}  2005, Dobbs-Dixon \&
Lin 2007).  We adopt a spherically symmetric approximation for the
surface temperature of the planet to be 1,360K. Here, we neglect the
modification in the internal structure due to the ohmic dissipation
which will be considered with self-consistent models in the next
sections.  In Paper IV, we will also consider the conductivity on the
planet's upper atmosphere due to photoionization which only occurs on
the dayside of the planet.

Using this conductivity profile, we can approximate the magnetic
diffusivity $\eta(r)=1/{\mu_{0}\sigma(r)}$ by:
\begin{equation}
\eta(r) \simeq 10^{3} {\rm exp}
\left[25\left(\dfrac{r}{R_{p}}\right)^{2}\right]
\label{expression of eta}
\end{equation}
where the effects of the photoionization have been neglected in this
paper.

To apply the procedure described in \S3, we also need to specify: 
\begin{flushleft}
\textbf{Relative angular velocity:} $\omega=10^{-5}\ s^{-1}$, \\
\textbf{Star's magnetic dipole:} $m=4\times 10^{34}\ A\ m^{2}$, \\
\textbf{Value of the tilt of the magnetic dipole:} ${\rm \sin}(\alpha)=1$. \\
\end{flushleft}
We then obtain the following $\mathcal{P}$ (also see figure 2 for the
graphs of $G_{l}(r)$)
\begin{equation}
\mathcal{P}(t)=2.26\times 10^{21}cos^{2}\omega t+2.1\times 
10^{21}\sin^{2}\omega t+1.3\times
10^{21} {\sin \omega t} {\cos \omega t} \label{W}
\end{equation}
We then take the average in time over one synodic period and obtain
$\mathcal{P}\approx 2.18 \times 10^{21}\ Js^{-1}$ \\

\begin{figure}[htbp]
\begin{center}
\epsscale{1.3}
\plottwo{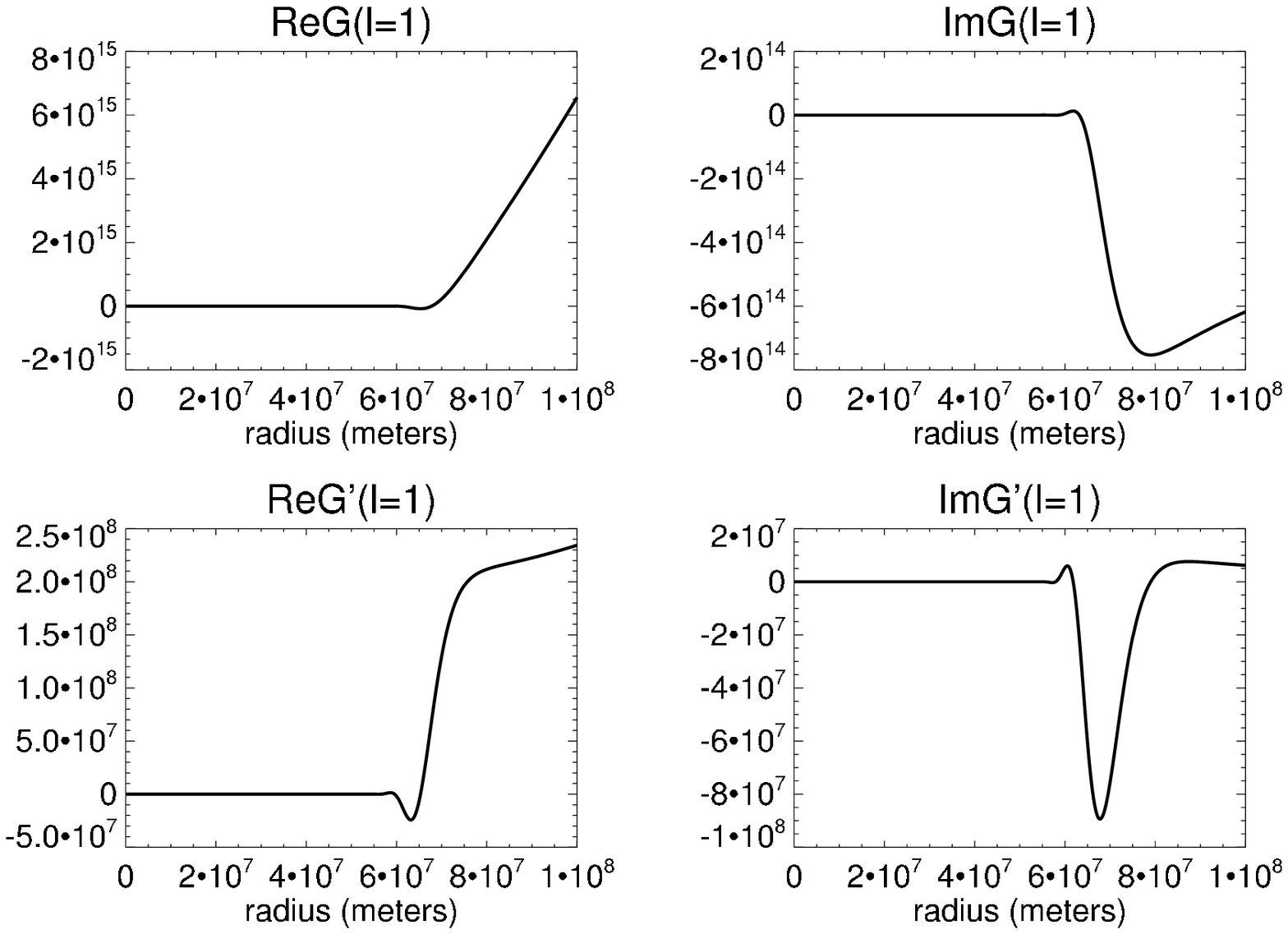}{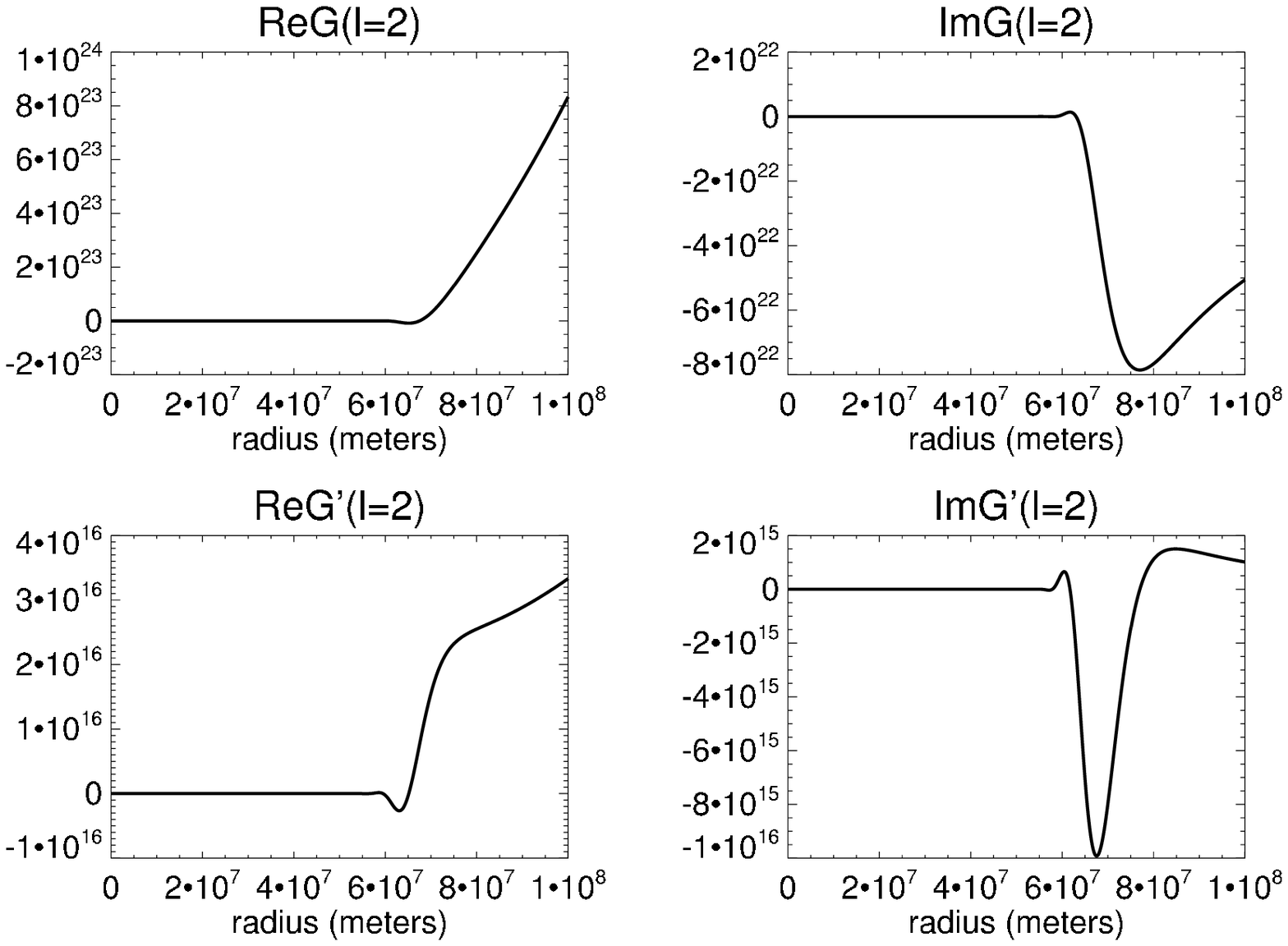}  
\label{Graph of G}
\end{center}
\caption{$Re(G_{l=1})(r)$, $Im(G_{l=1})(r)$, $Re(G_{l=2})(r)$,
$Im(G_{l=2})(r)$, and their first derivatives, for $R_{p}=10^{8}m=1.4\
R_{J}$, $a=0.04\ AU$, and $\eta(r) \simeq 10^{3} {\rm exp}
\left[25\left(\dfrac{r}{R_{p}}\right)^{2}\right]$ The shape of $G_{l}$
for $l=1$ and $l=2$ are very close but the amplitudes for $l=2$ are
about $10^{8}=R_{p}$ higher than for $l=1$.  Indeed, the major
difference between $l=1$ and $l=2$ is found in the equations
describing the boundary conduitions (see equation \ref{detailed
equations} where a factor $10^{8}$ between $l=1$ and $l=2$ comes from
the term $R_{p}^{l}$). In addition, we found $|C(l,m)|$ for $l=1$ is
about $10^{10}$ times larger than for $l=2$. Therefore $|G_{l}
C(l,m)|$ for $l=1$ is much larger than for $l=2$, which allows us to
keep only the terms corresponding to $l=1$ and $l=2$ in the
decomposition of $\phi$ on spherical harmonics.}
\end{figure}

The conductivity profile we have obtained here is sensitive to the
planetary structure model.  At the epoch of planet formation, the gas
accretion and planetesimal bombardment history are stochastic (Zhou \&
Lin 2007).  The opacity in the accretion envelope of proto gas giant
planets may also be subjected to variations due to dust coagulation
(Iaroslavitz {\it et al.} 2007).  The thermal evolution of these planets
can be highly diverse.  There may, therefore, be a dispersion in the
magnitude of $\eta$.

\subsection{Comments on the skin depth and the dependence of the ohmic 
dissipation on the conductivity and on the sign of $\omega$}
\label{sec:skin}
Once the conductivity profile within the planet is determined, we are
able to compute the energy dissipation rate inside the planet of the
current induced by the star's magnetic field.  In light of the
possible uncertainties in the magnitude of $\eta$, we compute the
ohmic dissipation rate for different $\eta$ by artificially modifying
the above determined $\eta$ with a multiplicative factor. The
resulting magnitude of the time-averaged value of $\mathcal{P}$ is
listed below (table \ref{artificially modified eta}).

\begin{table}[htbp]
\begin{center}
\begin{tabular}{|l|c|}
\hline
$\eta(r)=10^{-3}\ {\rm exp} (25\ (\dfrac{r}{R_{p}})^{2})$ & $\mathcal{P}=1.26\times 10^{21}\ Js^{-1}$  \\
\hline
$\eta(r)=10^{0} {\rm exp} (25\ (\dfrac{r}{R_{p}})^{2})$ & $\mathcal{P}=2.7\times 10^{21}\ Js^{-1}$ \\
\hline
$\eta(r)=10^{3}\ {\rm exp} (25\ (\dfrac{r}{R_{p}})^{2})$ & $\mathcal{P}=2.18\times 10^{21}\ Js^{-1}$ \\
\hline
$\eta(r)=10^{5}\ {\rm exp} (25\ (\dfrac{r}{R_{p}})^{2})$ & $\mathcal{P}=1.71\times 10^{21}\ Js^{-1}$ \\
\hline
$\eta(r)=10^{7}\ {\rm exp} (25\ (\dfrac{r}{R_{p}})^{2})$ & $\mathcal{P}=1.12\times 10^{21}\ Js^{-1}$ \\
\hline
$\eta(r)=10^{9}\ {\rm exp} (25\ (\dfrac{r}{R_{p}})^{2})$ & $\mathcal{P}=2.33\times 10^{20}\ Js^{-1}$ \\
\hline
$\eta(r)=10^{10}\ {\rm exp} (25\ (\dfrac{r}{R_{p}})^{2})$ & $\mathcal{P}=2.5\times 10^{19}\ Js^{-1}$ \\
\hline
$\eta(r)=10^{12}\ {\rm exp} (25\ (\dfrac{r}{R_{p}})^{2})$ & $\mathcal{P}=2.5\times 10^{17}\ Js^{-1}$  \\
\hline
\end{tabular}
\end{center}
\caption{Table giving $\mathcal{P}$ as a function of $\eta$, where the
value of $\eta$ is artificially modified from the value
computed in section \S4}
\label{artificially modified eta}
\end{table}

These results indicate that the energy dissipation rate is insensitive
to a change in the amplitude of the conductivity by several orders of
magnitude (this conclusion is in agreement with a conjecture that
Campbell made (C83)). A high conductivity increases the energy
dissipation in a given volume, but it also tends to prevent the
magnetic field from penetrating inside the planet. On the other hand,
a lower conductivity corresponds to less dissipation per unit of
volume, but it also allows the field to penetrate deeper inside the
planet (and therefore increasing the volume where energy can be
dissipated).

The skin depth (for reasonable values of $\eta(r)$) is of order
$\delta=\sqrt{\dfrac{\eta}{\omega}}$.  For $\eta(r)=10^{3}{\rm
exp}\left[25\left( \dfrac{x}{R_{p}} \right)^{2} \right]$ and
$\omega=10^{-5}s^{-1}$, we have $\delta(r_{pn}) \approx 4\times
10^{7}m$ (we define $r_{pn}$ to be the radius of penetration, or the
radius to which the magnetic field can diffuse inside the
planet). This estimate is consistent with the numerical values of
$G_{l}(r)$ inside the planet (see figure 2) in which
we find that $G_{1}(r)$ for $r< r_{pn} \simeq 6.5\times 10^{7}m$ is
negligibly small compared to its value elsewhere.

These considerations suggest that the total rate of energy dissipation
is well determined though the location where it occurs is less well
established due to the uncertainties in $\eta$. Moreover, with our
definition $\omega=\omega_{\star}-\Omega_{p}$, $\omega$ is positive
outside corrotation and negative inside corrotation. However the ohmic
dissipation rate inside the planet $\mathcal{P}$ only depends on the
absolute value of $\omega$.

\subsection{Energy source and direct influence on the planet's orbit.}

The induced current $\mathbf{J}$ deduced in
the previous section is due to the diffusion of a time dependent
magnetic field.  This time dependence comes from the relative motion
of the planet's orbit and the stellar magnetosphere. Thus, the ohmic
dissipation must be supplied by the orbital kinetic energy of the
planet and the rotational energy of the star.  Our stated goal in the
introduction is to consider whether the migration of some planets may
be halted by their magnetic coupling with their rapidly spinning
magnetized host stars.  In the case where $\omega_\ast > \Omega_p$,
the rotational energy of the star is transferred to the total orbital
energy of the planet and provides a supply for the ohmic
dissipation. The torque T associated with the ohmic dissipation  
is linked with the ohmic dissipation rate $\mathcal{P}$ and the relative angular velocity $\omega$ 
according to the following equation (cf. C83, eq. 55),
\begin{equation}
\mathcal{P}=-\omega |T|.     
\label{torque}
\end{equation}

Since the transfer of angular momentum requires the
torque associated with the ohmic dissipation, a similar fraction of
energy is being transfered to the planet's orbit and supplied to the
ohmic dissipation.  For this purpose, we qualitatively compare the
magnitude of $\mathcal{P}(t)$ with the rate of energy change needed
stall the migration of a protoplanet. A detailed computation on the
orbital evolution of the planet will be presented in Paper III.

For illustration purpose, we first consider the power associated with
the migration ($P_{mig}$) of a planet with a 0.63 Jupiter mass and a
1.4 Jupiter radius toward a sun-like star.  At any semi major axis
$a$, the total energy of the Keplerian orbit is $|E| = G M_p M_s/2a$.
If its orbit decays on a characteristic planet-disk interaction time
scale ($\tau_{\rm mig}$) of about 3 million years, the torque needed
to halt the planet's migration would correspond to a power $P_{mig}$
such that
\begin{center}
$P_{mig} = |\dot E| \simeq \dfrac{\mathcal{G}
M_{p} M_{\star}}{2\ a\ \tau} \simeq 7.4\ \times 10^{22} Js^{-1} $.
\label{migration kinetic power}
\end{center}
Since this power is more than an order of magnitude larger than the
time average value of $\mathcal{P}$ (see Table 1), it seems,
therefore, not possible for the magnetic coupling to directly stall
the planet's migration at a 0.04 AU Keplerian orbit within a few
millions years, even in the limit of a positive $\omega$.  

However in the model we have considered here, $\mathcal{P}\propto B^2
\propto m^2 a^{-6}$.  The power needed to drive the planet to
migration with a specified speed is proportional to $a^{-2}$ (these
scalings are confirmed by numerical calculations that neglect any
changes in the relative frequencies $\omega$ and the planetary
internal structure). It means that there is a semi-major axis
$a_{stop} (\sim 0.01$AU) at which $\mathcal{P}$ and $P_{mig}$ are
comparable.  This distance is comparable to the radius of a typical T
Tauri stars.  Note that the requirement for $\omega_\ast > \Omega_p$
also implies that the planet must be outside the corotation
radius. This condition is satisfied only in a disk with a low gas
accretion rate around a rapidly spinning and weakly magnetized
stars. In paper II, we will consider such a model for the newly
discovered planet around TW Hyd (Setiawan {\it et al.} 2008).  Under
these circumstances, the planet-star magnetic interaction may also
be overwhelmed by their tidal interaction.  

\section{Planetary inflation and mass loss}
In this section, we propose that ohmic dissipation in the planet's
interior can indirectly halt its migration.  The main physical
mechanisms involve the heating of the planet's interior, its inflation
and mass loss through Roche lobe overflow, and angular momentum
transfer from the transferred material to the orbit of the planet.

Up to now, we have computed the planet's conductivity for one
particular set of parameters ($M_{p}$, $a$, etc.), and the corresponding ohmic energy
dissipation inside the planet due to the star's magnetic field.
Although, this dissipation rate for most close-in planets is generally
too small to directly provide the power needed to halt their migration
over the time scale of a few Myr, it can modify their internal
structure.

The ohmic dissipation is likely to increase the temperature, the ionization
fraction, and the conductivity around the region where most of the
dissipation occurs.  In principle, the extra energy source would
reduce the skin depth. However, the envelope of the young planet is
likely to be fully convective, similar to the low-mass main sequence
secondary in interacting binaries.  Campbell (C83) suggested that the
dominant diffusivity may be due to turbulence (Cowling 1981).  In
\S\ref{sec:skin}, we have already indicated that even though the skin
depth may be affected by the magnitude of the diffusivity, the total
energy dissipation rate in the planet's interior is not sensitively
determined by the profile of $\eta$.

Nevertheless, the heat released by the dissipation is comparable to
that associated with the Kelvin-Helmholtz contraction during the early
stage of the planet's evolution (Bodenheimer, Lin, Mardling 2001 (BLM)).  In
the proximity of its host star, this extra energy source may cause a
planet to inflat beyond its Hill's radius and lose mass (Gu {\it et
al.} 2004).  

In the following sections (\S\S5-7), we adopt an idealized and
self-consistent model of the planet's internal structure.  This
approach allows us to compute the conductivity of the planet for
different sets of parameters. Considering the low dependence 
of the total ohmic dissipation $\mathcal{P}$ on $\eta$, an idealized but versatile
prescription is adequate for the computation of $\mathcal{P}$ and the
mass loss rate ($\dot M$) for different values of the important
parameters (the planet's mass and radius, the star's mass, luminosity,
and dipolar magnetic field strength, the tilt between the magnetic 
dipole and the stellar spinning axis, and the relative orbital period).  In
\S5, we show how the mass loss rate $\stackrel{\centerdot}{M}$ is
related to the ohmic dissipation $\mathcal{P}$.  In \S6, we describe the
model we used for the planet's interior and in \S7 we calculate $\mathcal{P}$
and $\stackrel{\centerdot}{M}$ for different sets of parameters.

\subsection{A qualitative description}
The planet receives energy, at its surface, from the star's radiation
and, in the interior, from the ohmic dissipation. The surface heating
diffuses inwards until an isothermal structure is established in the
planet's outer envelope. But, well below the surface region, the heat
flux is generated by the planet's Kelvin-Helmholtz contraction and
ohmic dissipation and transported by convection.  In the limit that
convection is efficient, the envelope attains a constant entropy
profile. For computational simplicity, we adopt an isothermal model
near the surface of the planet and a polytrope model for its deep
interior. 

There are two regions of interest. Very close to the host star, the
ohmic dissipation rate is larger than that ($L_p= L_\ast (R_p/2a)^2$)
due to the stellar irradiation ($L_\ast$) received by the planet. In
this limit, the planet would rapidly expand beyond its Roche lobe and
become tidally disrupted. In accordance with the results of the
previous section, (in which the effect of $\mathcal{P}$ on the
internal structure of the planet has been neglected), $\mathcal{P}
\propto a^{-6}$ and $L_p \propto a^{-2}$.  Thus, the stellar heating
dominates at larger semi major axis. In this section, we consider the
effect of planet's inflation due to the ohmic dissipation and show
that $\mathcal{P}$ also increases with the planetary radius $R_p$ at
nearly the same rate as $L_p (\propto R_p^2)$. Thus, during the
thermal expansion of the planet, the ratio of $L_p/\mathcal{P}$ does
not change.  In the region where $L_p > \mathcal{P}$, the effective
temperature at the planet's surface, with or without the contribution
from the ohmic dissipation remains to be the equilibrium value $T_p$.
But, the planet's radius for thermal equilibrium increases with
$\mathcal{P}$ which adds to the energy generation in the planet's
interior (BLM). If the new equilibrium $R_p$ is larger than the
planet's Roche radius, $R_H$, mass would be lost gradually through
Roche overflow.

\subsection{Mass loss rate} 
\label{mass loss rate section}
We now derive the equations that will allow us to calculate the mass
loss rate $\dot M$ and angular momentum transfer rate as functions of
$\mathcal{P}$. We are in the second region where the ohmic dissipation is less
than the radiation flux from the star ($\mathcal{P} \leqslant  L_{p}$), and we set the Bond albedo to zero. 
We therefore assume that the
equilibrium temperature at the surface of the planet is fixed by the
radiation from the star $T_{p}^{4}=\frac{L_{star}}{16\pi\ \sigma_{r}\
a^{2}}$ ($L_{star}$ the total luminosity of the star, $\sigma_{r}~=~
5.67\times 10^{-8}~Js^{-1}m^{-2}T^{-4}$), and that the ohmic dissipation
provides the additional energy to inflate the planet.

An irradiated short-period planet establishes an isothermal surface
layer. The hot interior continues to transport heat to this region and
then radiates to infinity with a luminosity $L_i$ despite the surface
heating. Note that
\begin{equation}
L_i < <  L_p = 4 \pi \sigma_{r} T_p^4 R_p^2
\end{equation}
so that the modification to $T_p$ is negligible.  The magnitude of
$L_i$ is a function of $R_p$, $M_p$, $T_p$, and the existence of the
core.  We have previous computed an equilibrium model for the
parameters for several short-period planets (BLM). 
In the range [$10^{-8} L_\odot 10^{-5} L_\odot$],
the numerical results of BLM can be approximated by 
\begin{equation}
{\rm log} {R_p \over R_\odot}  = A (M_p)
+ B(M_p) {\rm log} {L_i \over L_\odot} 
+ C(M_p) \left( {\rm log} {L_i \over L_\odot} \right)^2.
\label{eq:req}
\end{equation}
For HD209458b (the $0.63 M_J$
model we presented in the previous section), (A, B, C)= (3.11, 1.01,
0.0642).  BLM also determined the value of these coefficients for more
massive planets around a solar type star (they are modified by the
stellar irradiation so that they are also function of $M_\star$).  The
planet's radius $R_p$ would contract unless there is an adequate energy
source to replenish its loss of internal energy. If the ohmic
dissipation can provide such a source, $R_p=R_e$ and $L_i = \mathcal{P}$ in a
thermal equilibrium.

At a=0.04 AU, the Roche radius of the planet is 
\begin{equation}
R_H=a\left(\dfrac{M_{p}}{3M_{\star}}\right)^{\frac{1}{3}}.
\label{definition hill's radius}
\end{equation}
From equation(\ref{eq:req}), we find that the equilibrium $R_e \sim
R_H$ if $L_i \sim 10^{-5} L_\odot$ which is approximately the value of
$\mathcal{P} (\sim 10^{21}$ J s$^{-1})$ we have determined for HD209458b.  During
the planet and star's infancy, this planet would inflate to fill its
Roche lobe when it has migrated to this location.

Outside $\sim 0.04$ AU, $\mathcal{P}$ decreases rapidly with
$a$. Consequently, $R_p$ reduces to the value which is essentially not
modified by the ohmic heating. For the calculation presented in the
previous section, we neglected the inflation of the planet. In the
next section, we construct a self consistent model taking into account
of the modification of the dissipation rate due to the internal
structural changes.  For $a=0.04$ AU, the intense ohmic dissipation
rate (with $L_i = \mathcal{P} \simeq 2 \times 10^{21}$J s$^{-1}$)
modifies the planet's internal structure and inflates its radius to
$R_e \sim 0.5 R_\odot$.  The inflation is more severe at $a < 0.04$ AU
because $\mathcal{P}$ is a rapidly decreasing function of $a$.  If
$R_e >R_H$ at this location, the planet would overflow its Roche lobe
and loss mass.  For the rest of this paper, we assume that the we are
in the case where the planet fills its Roche lobe, \textit{i.e.}
$R_{p}=R_{H}$

Two remarks are appropriate here. Firstly, in order for the Roche lobe
overflow to provide angular momentum, the actual shape of the Roche
lobe should be taken into account. However, for computational
simplicity, we adopt in this paper a spherically symmetric
approximation (refer to Gu, Bodenheimer, Lin 2003 for a detailed study
(GBL)). Secondly, we have only considered the contribution of ohmic
dissipation $\mathcal{P}$ to the planetary inflation. The tidal
(gravitational) interaction between the star and the planet can also
significantly enhance the planet's inflation in some cases. More
precisely, this tidal interaction can be strong for small semi-major
axis (the tidal effect varies as $a^{-13/2}$) and for large radius
(thus, the more inflated the planet, the stronger this effect
becomes).

\subsection{The governing equations}
Mass loss process is initiated when $R_{e} \geq R_{H}$. 
In this limit, a contiuous flow would be
established in which the inflation of the envelope's drives a steady
supply of gas to the Roche lobe region.  Well inside the Roche lobe,
the gravitational potential is primarily determined by the mass of the
planet $M_{p}$: $\phi_{g}=-\frac{GM_{p}}{r}$, but near $R_H$, we need
to take into account of both the planet and the star.  In a frame which 
corotates with the planet, the gravitational potential U(r) 
\begin{equation}
U(r)=\dfrac{-GM_{\star}}{a}\left[\left(1-\dfrac{M_{p}}{M_{\star}}\right)\ 
\left(\dfrac{a}{a-r}+\dfrac{(a-r)^{2}}{2a^{2}}\right)+\dfrac{M_{p}}
{M_{\star}}\left(\dfrac{a}{r}+\dfrac{r^{2}}{2a^{2}}\right)\right]
\label{expression of U}
\end{equation}
where $r$ is the distance to the center of the planet. The value of
$R_H$ is determined from ${dU}/{dr}(R_H)=0$.

In principle, this potential introduces a complex multi-dimensional
flow pattern, especially near the Roche lobe.  But the expansion of
the envelope originates deep in the envelope where the ohmic
dissipation occurs.  In this region, spherical symmetry is adequate.
Near the Roche lobe, we adopt the results obtained by GBL. 
For computational convenience, we neglect the planet's spin.

We consider a low-velocity quasi-hydrostatic expansion of the envelope.
Under this gravitational potential in eq. (\ref{expression of U}), the
radial component of the hydrodynamics momentum equation for a volume
of gas is reduced to:
\begin{equation}
\upsilon\dfrac{d\upsilon}{dr}(r)+\dfrac{1}{\rho(r)}\dfrac{dP}{dr}(r)
=-\dfrac{1}{\rho(r)}\dfrac{dU}{dr}(r)
\label{momentum 1}
\end{equation}
where $v$ is the radial velocity.  The radial component of the
equation of mass conservation: $\nabla (\rho (r) \upsilon(r))=0$ gives
$r^{2}\rho (r)\upsilon(r)=Constant$, and then the mass loss rate
$\stackrel{\centerdot}{M}$ is constant:
\begin{equation}
\stackrel{\centerdot}{M}=4\pi\ r^{2}\rho (r)\upsilon(r)=Constant
\label{definition mass loss rate}
\end{equation}
or equivalently,
$\dfrac{1}{\rho}\dfrac{d\rho}{dr}=-\dfrac{1}{r^{2}\upsilon}
\dfrac{dr^{2}\upsilon}{dr}$. Then using $\dfrac{dP}{dr} =\dfrac{dP}
{d\rho}\dfrac{d\rho}{dr} =\dfrac{d\rho}{dr}c_{s}^{2}$ ($c_{s}^{2}(r)$
is the sound speed), the momentum equation becomes:
\begin{eqnarray}
\left(1-\dfrac{c_{s}^{2}}{\upsilon^{2}}\right)\dfrac{d}{dr}
\left(\dfrac{\upsilon^{2}}{2}\right)&=&-\dfrac{1}{\rho}
\dfrac{dU}{dr}\left(1-\dfrac{2c_{s}^{2}\rho}{r}
\dfrac{1}{\frac{dU}{dr}}\right) .
\label{momentum equation}
\end{eqnarray}
At a (sonic) radius $r_{2}$ near the inner Langragian point, the flow
velocity becomes comparable to the sound speed (GBL) {\i.e.}
\begin{equation}
\upsilon(r_{2})=c_{s}(r_{2}).
\end{equation}
where the magnitude of $r_{2}$ is the largest solution of
$(r^{2}-rR_{H}+\frac{2c_{s}^{2}a^{3}}{LGM_{\star}})=0$, with
\begin{equation} 
L=\left(1-\dfrac{M_{p}}{M_{\star}}\right)\ \left(\dfrac{2a^{3}}{(a-R_{H})^{3}}+1\right) 
+ \dfrac{M_{p}}{M_{\star}}\left(2\left(\dfrac{a}{R_{H}}\right)^{3}+1\right)\simeq 9 \label{l}.
\end{equation}

The expansion rate is determined by the rate of ohmic energy
dissipation within the planet. In a steady state, the energy equation
reduces to
\begin{equation}
\dfrac{1}{r^{2}}\dfrac{d}{dr}\left[r^{2}\rho(r)\upsilon(r)\left(
\dfrac{\upsilon^{2}}{2}+h(r)+\phi_{g}(r)\right)\right]=\mathcal{P}_{vol}
\label{energy equation}
\end{equation}
where $\mathcal{P}_{vol}$ is the volumic ohmic energy dissipation, and
$\phi_{g}$ the gravitational potential (of the planet only or of both
the planet and the star, depending on the location).  In this
approximation, we assume that the distribution of enthalpy $h$ is
determined by both efficient convective transport (in term of an
abiabat) and radiative diffusion inside the planet.

Using equation (\ref{definition mass loss rate}), we replace 
$r^{2}\rho(r)\upsilon(r)$ by $\dfrac{\stackrel{\centerdot}{M}}{4\pi}$ 
in eq. \ref{energy equation}. We then can then integrate
equation (\ref{energy equation}) between the radius $r_{pn}$
(the radius at which the field can no longer penetrate into the
planet) and $r_{2}$ so that
\begin{equation}
\stackrel{\centerdot}{M}\left(\dfrac{\upsilon^{2}}{2}(r_{2})-
\dfrac{\upsilon^{2}}{2}(r_{pn})+h(r_{2})-h(r_{pn})
+\phi_{g}(r_{2})-\phi_{g}(r_{pn})\right)=\int
4\pi r^{2}\mathcal{P}_{vol}(r)dr=\mathcal{P}.
\label{eq:eofw}
\end{equation}
Within an order of magnitude, $\bigtriangleup(h)\approx-\dfrac{1}{3}
\bigtriangleup(\phi_{g})$ and $\bigtriangleup(\dfrac{1}{2}v^{2})
\approx\dfrac{1}{10} \bigtriangleup(\phi_{g})$ (this comes from calculating the order of magnitude of these 3 terms using an 
order of magnitude for the temperature, for the sound speed, and for $r_{pn}$). In addition, the
integrated energy equation is the result of an approximation as the
total ohmic dissipation rate $\mathcal{P}$ should be the integral of
$\mathcal{P}_{vol}$ between $r_{pn}$, and $R_{p}$ ($=R_{H}$ because we
assumed that the planet fills its Roche lobe).  However, this
approximation is reasonable since $h(R_{H}) =h(r_{2})$ (the surface
region is approximately isothermal), $\upsilon^{2}(r_{2})\approx
\upsilon^{2}(R_{H})$, and $\phi_{g}(r_{2})\approx \phi_{g}(R_{H}$
because $r_{2}$ is very close to $R_{H}$).

We now can calculate $\rho(r_{2})$ and $P(r_{2})$. Equation
(\ref{definition mass loss rate}) for $r=r_{2}$ with
$\upsilon(r_{2})=c_{s}(r_{2})=10^{4}\sqrt{\dfrac{T}{10^{4}}}$ meters
gives:
\begin{eqnarray}
\left\{
\begin{array}{l}
\rho(r_{2})=\dfrac{\stackrel{\centerdot}{M}}{4\pi r_{2}^{2}c_{s}(r_{2})}\\
P(r_{2})=\alpha\ \rho(r_{2}) T(r_{2})\\
\alpha=\dfrac{\mathcal{N}_{a}k_{B}}{\mu\mathcal{M}_{H}}\
\label{values of P and volumic mass at r2}
\end{array}
\right.
\label{P(r2)}
\end{eqnarray}
where $\mathcal{N}_{a}$ is the Avogadro constant, $k_{B}$ the
Boltzmann constant, $\mathcal{M}_{H}$ is the hydrogen molar mass, 
$\mu$ a coefficient which depend on the ionisation
rate ($\mu=1$ for hydrogen atoms, and
$\mu=0.5$ for fully ionized hydrogen gas, and we usually choose $\mu$ close to unity).

\section{Isothermal and polytropic model} 
\label{isothermal and polytropic model}

In \S4, the permeation and dissipation of the time-dependent external
field is analyzed by neglecting any resulting changes in the planet's
interior.  In \S5, we show that the resulting ohmic dissipation can
substantially modify the temperature and density distribution within
the planet.  Increases in the ionization rate modify the skin depth
and relocate the region of maximum ohmic dissipation. However, the
expansion of the planet's envelope does not affect the rate of ohmic
dissipation.  In this section, we present a set of approximately
self-consistent calculations to analyze the feedback effect of ohmic
dissipation on the planet's internal structure.

\subsection{A Roche-lobe filling structural Model}
In principle, the structure of the planet should be solved numerically
with the standard planetary structure equations (BLM).  However, a
semi analytic model based on simplifying assumption may provide
insight on the inter-dependent relation between various physical
parameters.  Based on the BLM's numerical models, we approximate the
internal structure of the planet with an idealized model in which the
outer region is isothermal (due to the stellar irradiation) and the
inner region is polytropic (due to an efficient mix of entropy by
thermal convection).  In the computation of $\eta$, we only take into
account the ionization of the hydrogen because the internal
temperature distribution is mostly determined by heat transfer rather
than heat dissipation and the rate of $\mathcal{P}$ is a relatively
insensitive function of $\eta$. The advantage of this approximation is
that its application for the self consistent analysis is relatively
straightforward.

\begin{description}
\item[The isothermal region:] it extends from the surface to a
transition radius $r_{=}$ which is to be determined self consistently
in \S \ref{subsection transition between the two models}.  In this
region, the equation of state and the equation describing the
hydrostatic equilibrium are:
\begin{eqnarray}
T(r)&=& {\rm Constant}, 
\label{isothermal T} 
\\ P(r)&=&\alpha\ \rho (r)\ T(r), \\ 
\label{equation of state}
\dfrac{dP}{dr}(r)&=&-\dfrac{GM_{int}(r)}{r^{2}}\rho (r), \
\label{hydrostatic equilibrium}
\end{eqnarray}
where $\alpha=\dfrac{\mathcal{N}_{a}k_{B}}{\mu\mathcal{M}_{H}}$ and
$M_{int}(r)$ is the planet's mass inside a sphere of radius r centered
on the planet's center. For all practical purpose, $\rho$ is
sufficiently low in the isothermal region that we can approximate
$M_{int}(r) \simeq M_{planet}$ (one can verify, a posteriori, that the
neglected mass is less than a few percent of the total mass). To
calculate P(r), we integrate (\ref{hydrostatic equilibrium}) using
(\ref{isothermal T}) and (\ref{equation of state}). We then can
calculate $\rho(r)$ using (\ref{equation of state}):
\begin{eqnarray}
\left\{
\begin{array}{l}
P(r)=C\ {\rm exp} \left(\dfrac{GM_{planet}}{r}\ \dfrac{1}{\alpha
T}\right) \\ 
\label{P isothermal model} 
\rho(r)=\dfrac{1}{\alpha T}\ C\ {\rm exp}\left(\dfrac{GM_{planet}}{r}\
\dfrac{1}{\alpha T}\right)
\label{volumic mass isothermal model}
\end{array}
\right.
\label{P in isothermal region}
\end{eqnarray}
where $C$ is an integration constant, which value is obtained 
by injecting $r_{2}$ in the previous equations. 

\item[Polytrope region:] it extends from the center of the planet to
$r_{=}$.  In this region, we use the following equations:  
\begin{eqnarray}
P(r)&=&K\rho^{\gamma}(r), \label{polytrope equation of state} \\
\dfrac{d\phi_{g}}{dr}(r)&=&\dfrac{G\ M_{int}(r)}{r^{2}}, \label{definition gravitational potential}\\
\dfrac{dP}{dr}(r)&=&-\dfrac{GM_{int}(r)}{r^{2}}\rho (r) \label{polytrope hydrostatic equilibrium}\\
\bigtriangleup\phi_{g}(r)&=&4\pi\ G\ \rho(r) \label{poisson equation}
\end{eqnarray}
where $\phi_{g}$ is the gravitational potential, and $\bigtriangleup$ 
is the laplacian (in the Poisson equation). 
Using equations (\ref{polytrope equation of state}) and (\ref{definition
gravitational potential}), equation (\ref{polytrope hydrostatic equilibrium})
becomes: $K\gamma \rho^{\gamma
-1}(r)\dfrac{d\rho}{dr}(r)=-\rho(r) \dfrac{d\phi_{g}}{dr}$\\ And
after integration:\ \ \ \ \ \ $\phi_{g}(r)=Constant\ -\
\dfrac{K\gamma}{\gamma-1}\rho^{\gamma-1}(r)$\\ We then replace
$\phi_{g}$ in the poisson equation (\ref{poisson equation}):
\begin{equation}
\bigtriangleup \rho^{\gamma-1}(r)=-\frac{\gamma -1}{K\ \gamma}\ 
4\pi G\rho(r)
\end{equation}
For the condition appropriate in the interior of planets, the equation
of state is reasonably approximated by a $\gamma=2$ polytrope (de
Pater \& Lissauer 2001). In spherical coordinates, the previous
equation becomes:
\begin{equation}
\dfrac{1}{r^{2}}\dfrac{d}{dr}\left(r^{2}\dfrac{d}{dr}\rho(r)\right)
=-\dfrac{2\pi G}{K}\rho(r) .
\label{final poisson equation}
\end{equation}
This equation has an analytical solution (Ogilvie \& Lin
2004), and we can calculate $\rho(r)$, P(r) and T(r) in the region
described by the polytropic equation of state:
\begin{eqnarray}
\left\{
\begin{array}{l}
\rho(r)=\rho_{0}\dfrac{\sin kr}{kr} \\ \label{volumic mass
polytropic model}
P(r)=K\rho^{2}(r)=K\rho_{0}^{2}\left(\dfrac{\sin kr}{kr}\right)^{2} \\
\label{P polytropic model} 
T(r)=\dfrac{P}{\alpha \rho}(r)=\dfrac{1}{\alpha} K\rho_{0}\dfrac{\sin kr}{kr}
\\ \label{T polytropic model} k=\sqrt{\dfrac{2\pi G}{K}}.
\label{relation k and K}
\end{array}
\right.
\end{eqnarray}
\end{description}

\subsection{Transition between the two models}
\label{subsection transition between the two models}
In principle, the transition between the two regions is determined by
the onset of convection.  In the construction of hydrostatic
equilibrium structure models (to be presented in Paper II), we will
indeed use that condition to determine its photospheric radius.
Qualitatively, we expect the transition radius which separates the two
regions, $r_=$ to be larger than $r_{pn}$, because only in the
region interior to $r_=$ do we expect the temperature, ionization
fraction, and conductivity to be sufficiently large to halt the
penetration of the field.  In a hydrostatic equilibrium, the actual
value of $r_=$ is determined by the ratio of the ohmic dissipation
rate in the convective region to the sum of the ohmic dissipation rate
in the entire planet's interior and the stellar irradiative flux on
the planet's surface.  A set of fully self-consistent solution
requires the matching of the ohmic dissipation rate to be expected
from the planetary structure and that which determines its density and
temperature distribution (see paper II).

In the present context, we are considering the situation in which the
planet's radius is constrained by its Roche lobe and the density and
temperature of the outer boundary is determined by 
equation(\ref{P(r2)}). In this configuration, heat is also transported
by advection which modifies the location of $r_=$.  Moreover, the
density ratio between the planet's center and the outer boundary is
much larger than the temperature ratio.  Therefore, the polytropic
region cannot fill the entire interior region.  Since the pressure
scale height on the planet's surface is much smaller than its radius,
the isothermal region also cannot occur in the entire planet's interior
while containing all of its mass. Instead, the planet's interior
adjusts to attain a balance between the requirement of mass loading
and constraints set by hydrostatic equilibrium for appropriate
equations of state.  
 
In order to construct such an equilibrium model, we now determine
$\rho_{0}$ and k at $r_{=}$ where the transition between the two
regions occur. There are three equations that constrain these parameters:
$T_{isothermal}(r_{=}) =T_{polytropic}(r_{=})$, $P_{isothermal}(r_{=})
=P_{polytropic}(r_{=})$, and the total mass is constant. The first two
conditions also imply
$\rho_{isothermal}(r_{=})=\rho_{polytropic}(r_{=})$.  Therefore, we
solve the following equations for $\rho_{0}$, k, and $r_{=}$:
\begin{eqnarray}
\left\{
\begin{array}{l}
k^{2}=\dfrac{2\pi G\ C}{(\alpha T)^{2}} {\rm exp}\left(
\dfrac{GM_{\rm planet}}{\alpha T\ r}\right)\\
\rho_{0}=\dfrac{\alpha T}{2\pi G}\ k^{2}\
\dfrac{kr_{=}}{\sin kr_{=}} \\ \int_0^a 4\pi
r^{2}\rho_{\rm polytropic}(r)dr\ +\ \int_a^R 4\pi
r^{2}\rho_{\rm isothermal}(r)dr=M_{p}.
\end{array}
\right.
\label{3 unknowns}
\end{eqnarray}

By assuming an isothermal structure in the outer envelope,
we have neglected an outward heat flux.  This approximation is
only adequate if the dissipation rate is above that which is
need to inflat $R_{p}$ to the planet's Roche radius.  If this
condition is not satisfied, the planet's radius would attain
equilibrium values for which the surface cooling is balanced by the
Ohmic dissipation and stellar irradiation. We will construct, in Paper II,
the equivalent of equation 21 (for a 0.63 $M_{\odot}$ planet) which takes into
account the effect of ohmic dissipation in the planetary interior.

Whereas the temperature on the planet's surface is determined by the
stellar irradiation, the density at $r_2=R_H$ is determined by the
magnitude of $\dot M$ (through equation \ref{P(r2)}) which in term is
determined by the rate of ohmic energy dissipation $\mathcal{P}$ (see \S
\ref{sec:selfcon}).

For very large values of $\mathcal{P}$, a set of fully self-consistent solutions
also modifies the temperature at the disk surface as well as the
thermal content of the outflowing gas. However, provided $\mathcal{P}$ is small
compared with the stellar irradiative flux, a transition for
convective stability occurs near $r_=$.  

\subsection{Calculation of the magnetic diffusivity}
With these internal structure specified, we consider Saha's equation
for the hydrogen atoms which gives the ionization
fraction $x$ (Kippenhahn \& Weigertal, pages 107-111):
\begin{equation}
\dfrac{x^{2}}{1-x^{2}}=K_{H}=\dfrac{1}{P(r)}\dfrac{(2\pi m_{e})^{
\frac{3}{2}}}{h^{3}}(kT)^{\frac{5}{2}}exp\left(-\frac{E}{kT}\right) \ \ \ \ \ \ \ 
 \label{kh}
\end{equation}
where the ionization energy of hydrogen is $E=13.6eV$. We also
neglect here the radiation pressure as we write $P_{gas}(r)=P(r)$.

If the ionization fraction $x$ is small, $x^{2}\approx K_{H}$ (this is typically the case 
in the region where the ohmic dissipation occurs).  We use
$\sigma=\dfrac{N_{e}e^{2}}{m_{e}\nu_{e}}$, with 
$\nu_{e}=N_{n}10^{-19}(\dfrac{128kT}{9\pi m_{e}})^{\dfrac{1}{2}}$.  

The electric conductivity we would obtain 
does not take into account higher ionization states or the ionization 
of elements other than hydrogen atoms. We then use for the following 
calculations an electric conductivity that is 10 times higher than 
that we would obtain with the Saha equation (eq. (\ref{kh})) for the hydrogen atom only. 
We saw in table 1 that the ohmic dissipation rate was quite insensitive to the 
magnetic diffusivity $\eta(r)=(\mu_{0}\sigma(r))$, and we verified that this is also the case with 
the internal model we used for the planet in sections 5 to 8  
(for example, in this model, a uniform change in the magnetic diffusivity by a factor 10 changes 
$\mathcal{P}$ and $\stackrel{\centerdot}{M}$ by less than 20\%, and  
a uniform change in the magnetic diffusivity by a factor 100 changes 
$\mathcal{P}$ and $\stackrel{\centerdot}{M}$ by less than 40\%).

We then obtain the following expression for the magnetic diffusivity inside
the planet: 
\begin{equation}
\eta(r)=1.28\times 10^{-2}\frac{\sqrt{P(r)}}{T^{\frac{3}{4}}(r)}exp\left(\frac{78909}{T}\right). \label{eta}
\end{equation}
where T(r) and P(r) are the temperature and pressure of
the isothermal or polytropic region, depending on the radius r.

\section{Ohmic dissipation rate and the mass loss rate for 
different sets of parameters} 
\label{sec:selfcon}
With the above idealized prescription for the planet's internal
structure, we now calcultate self-consistently the ohmic dissipation
rate $\mathcal{P}$ inside the planet as well as the mass loss rate
$\stackrel{\centerdot}{M}$.

\subsection{Parameters involved in the calculation}
The model parameters involved in the calculation of the ohmic
dissipation rate inside the planet are: 1) the planet's mass $M_{p}$,
2) semi-major axis $a$, 3) the relative angular velocity $\omega$ (the
angular velocity of the field seen in a frame centered on the star and
rotating with the planet), 4) the strength of the star's magnetic
dipole moment $m$, and 5) the angle $\alpha$ between the spin axis of
the star and the star's magnetic dipole.

We use the isothermal and polytropic prescription described in the
previous section to model the planet's internal structure and
calculate the conductivity profile inside the planet. To do so, we
also need to specify 6) the mass of the star $M_{\star}$, and 7) the
star's luminosity $\L_{\star}$.

\subsection{Methodology}
The construction of a self-consistent model requires a loop of
retroaction involving the determination of the internal structure of
the planet and that of the ohmic energy dissipation.  For a specified
internal structure of the planet, one can compute (following \S3) the
conductivity profile and then the total ohmic dissipation rate
$\mathcal{P}$. However, this energy dissipated inside the planet
corresponds to an input of heat, which triggers an adjustment in the
planet's internal parameters. Due to the efficient convection inside
the planet, we assume that the adjustment of the internal parameters
to this external heating is quick.  We consider that the
characteristic time scale for the planet to evolve from one
equilibrium state to another is small compared to the variation time
scale of the seven parameters mentioned in the previous
paragraph. Therefore, we do not need to follow the planet's dynamical
evolution at all times. Instead, we can take a series of "snap shots"
of the planet in its equilibrium state for different set of
parameters.

Because of this feedback loop between the ohmic dissipation rate and
the planet's internal parameters, we use an iterative method. For any
chosen set of parameters, we start from an estimate for the ohmic
dissipation rate $\mathcal{P}_{0}$ and internal structure $T_{0}(r)$,
$P_{0}(r)$, and $\rho_{0}(r)$ corresponding to a magnetic diffusivity
profile $\eta_{0}(r)$ (In our parametric analyses, we typically make
small incremental changes in the model parameters from those for which
we have already obtained equilibrium values).  We then compute the new
internal structure $T_{1}(r)$, $P_{1}(r)$, and $\rho_{1}(r)$
associated with $\mathcal{P}_{0}$. This enables us to compute the
corresponding magnetic diffusivity $\eta_{1}(r)$. Finally, we use
$\eta_{1}(r)$ to calculate the corresponding ohmic dissipation rate
$\mathcal{P}_{1}$ and mass loss rate $\stackrel{\centerdot}{M_{1}}$.
This process is iterated until convergence of $\mathcal{P}$,
$\stackrel{\centerdot}{M}$, and of the internal parameters. Moreover,
for some specific set of parameters $M_{p}$, $a$, $\omega$, $m$,
$\sin(\alpha)$, $M_{\star}$, $L_{\star}$), we have started the
iterative process from two different initial states in order to verify
that they both converge to the same solution. Therefore, the iterative
process does converge to a unique solution.

We consider the following fiducial model in which the mass of the planet and semi-major axis corresponds to 
HD 209458 b, and in which the other parameters are reasonable ones for the type of systems considered. 
An estimate of the order of magnitude for the strength of the magnetic dipole can be found in Johns-Krull 2007. 
\begin{flushleft}
\textbf{Mass of the planet:} $M_{p}=0.63\ M_{J}=1.26\times 10^{27}\
kg$, \\ \textbf{Semi-major axis:} $a=0.04\ AU=6\times 10^{9}\ m$, \\
\textbf{Relative angular velocity:} $\omega=10^{-5}\ s^{-1}$, \\
\textbf{Star's magnetic dipole:} $m=4\times 10^{34}\ A\ m^{2}$, \\
\textbf{Value of the tilt of the magnetic dipole:} $\sin(\alpha)=1$,
\\ \textbf{Mass of the star:} $M_{\star}=M_{\odot}=2\times 10^{30}\
kg$, \\ \textbf{Luminosity of the star:} $\L_{\star}=1.5\
L_{\odot}=5.7\times 10^{26}$ W. \\
\end{flushleft}

\subsection{Computation of $\mathcal{P}$ and $\stackrel{\centerdot}{M}$, 
plots, and mathematical relations}

We present seven groups of plots (figures 3, 4, 5, 6, 7, 8, 9), one
group for each parameter mentioned just above. For each group, we vary
one parameter (x-axis), while keeping the others at the reference
values mentioned above. On the y-axis, we plotted the ohmic
dissipation rate $\mathcal{P}$, mass loss rate
$\stackrel{\centerdot}{M}$, and characteristic time scale $\tau_M =
\dfrac{M}{\stackrel{\centerdot}{M}}$. Note that the magnitude of 
$\tau_M \sim$ for a Jupiter mass planet is about 1Myr. In addition, 
the mass loss rate determined here is many orders of magnitude larger than
that due to photo evaporation.  Only with such large mass loss rate, can we
compensate for the angular momentum transfer due to the planet-disk
and planet-star tidal interaction.

We emphasize once again that in the construction of these models, we
assume that there is adequate energy dissipation to inflate the planet
with $R_e > R_H$. In later papers that use the roche-filling model, we verify
that $\mathcal{P} (R_H) > L_i (R_{H})$ before these results are
applied.  If this condition is not satisfied, the planet would not
fill its Roche lobe and not lose mass.

\begin{figure}[hbt]
\begin{center}
\epsscale{0.65}
\plotone{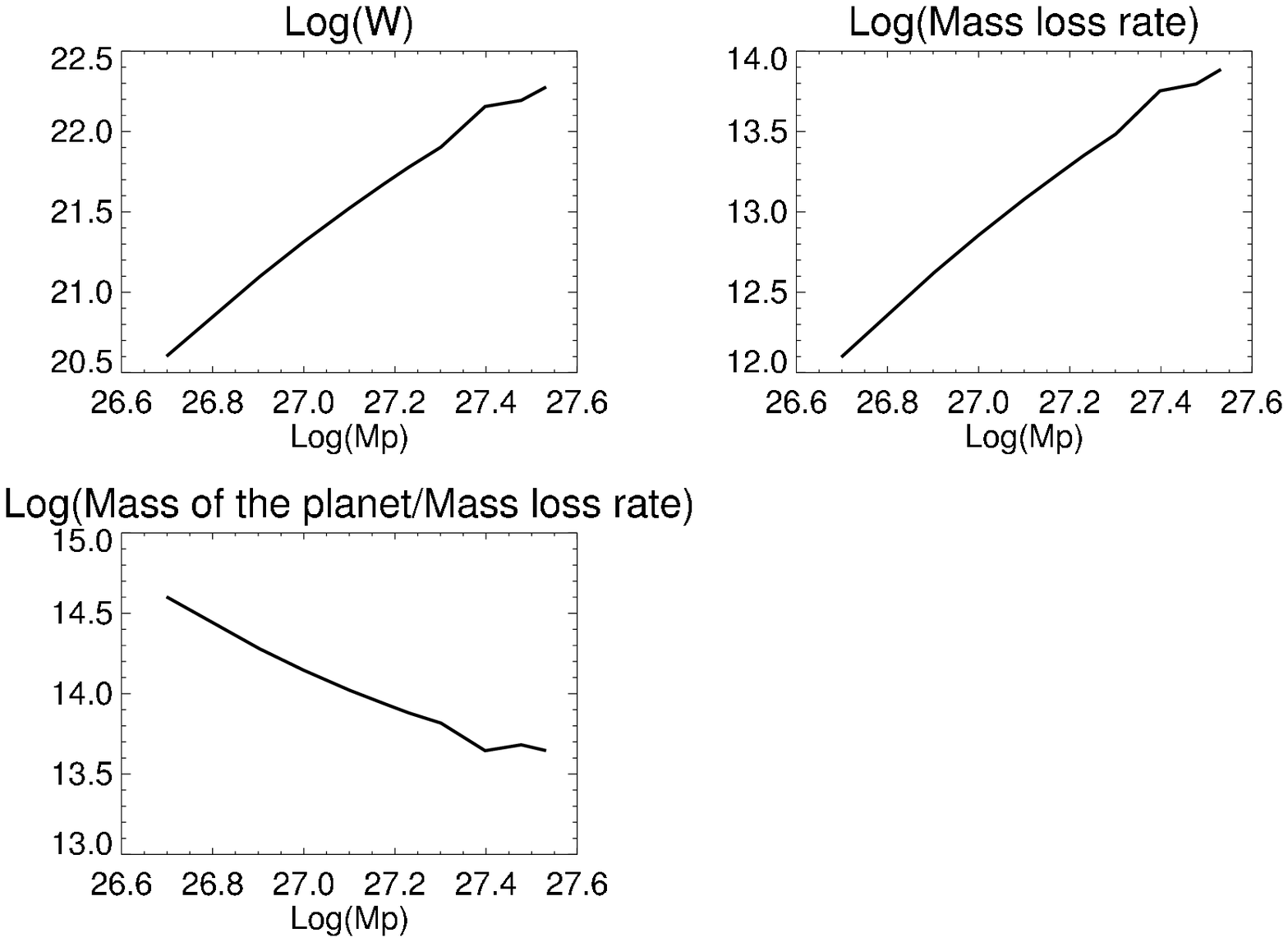}  
\end{center}
\label{Varying mass of planet}
\caption{Ohmic dissipation rate, mass loss rate, and time scale for
different planetary masses}
\end{figure} 

\begin{figure}[hbt]
\begin{center}
\epsscale{0.65}
\plotone{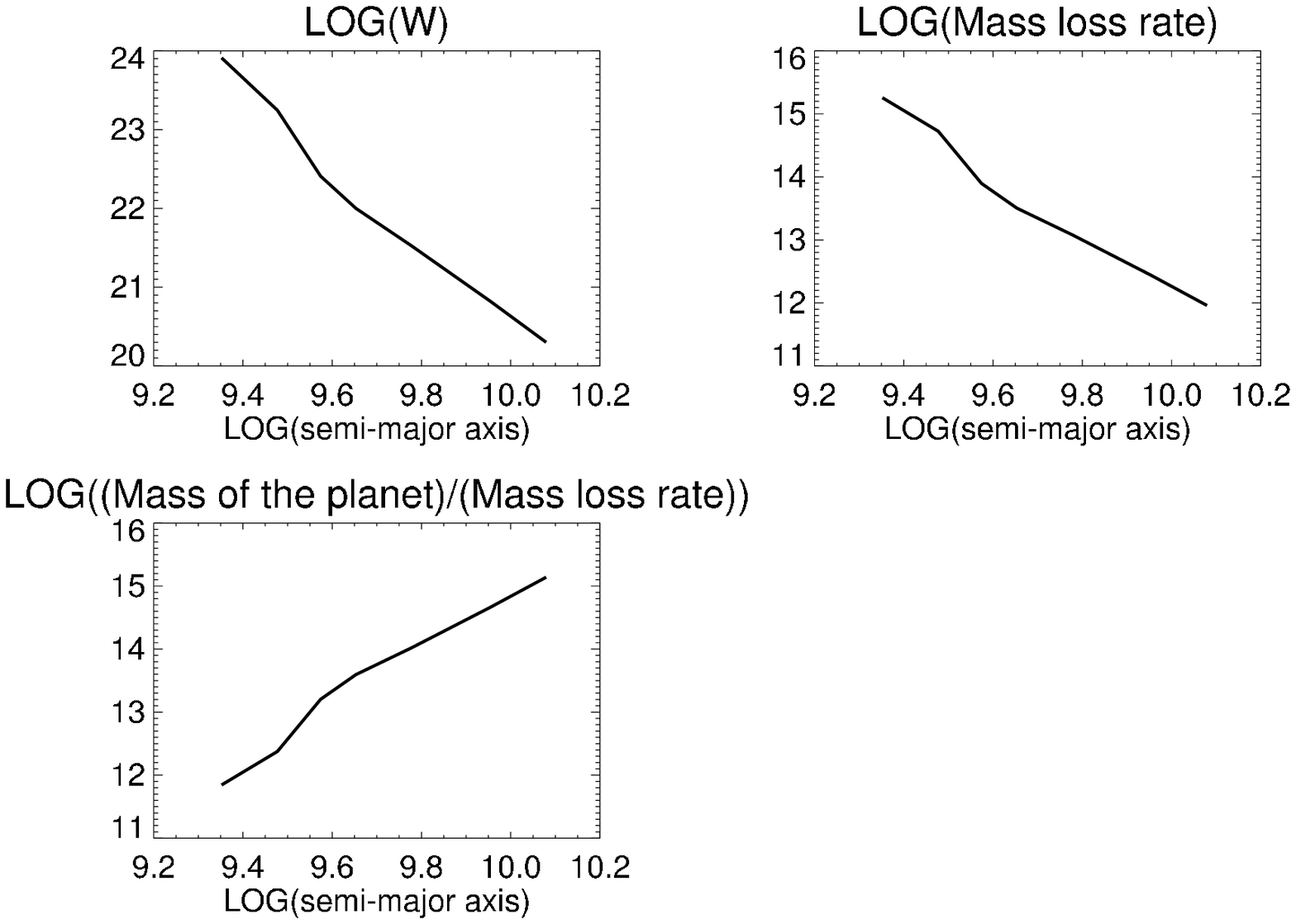}  
\label{Varying semi-major axis of the planet}
\end{center}
\caption{Ohmic dissipation rate, mass loss rate, and 
time scale for different semi-major axies}
\end{figure} 

\begin{figure}[hbt]
\begin{center}
\epsscale{0.65}
\plotone{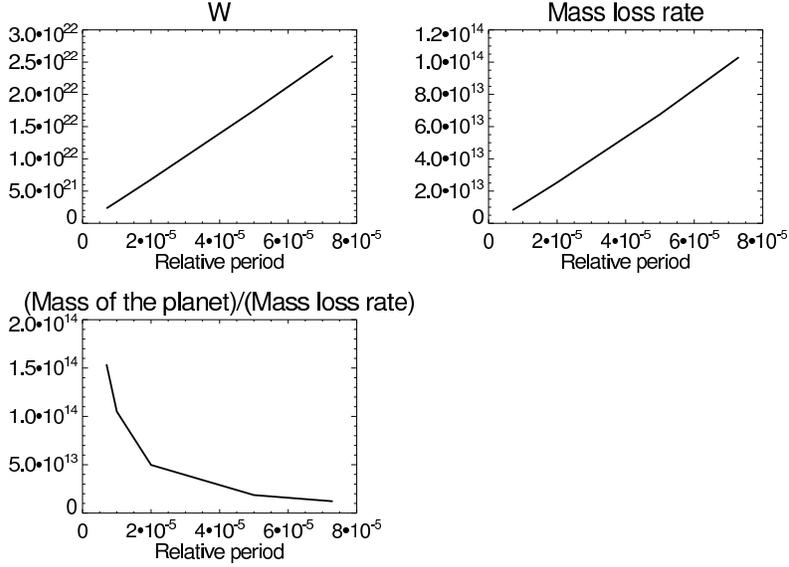}  
\label{Varying the relative angular velocity}
\end{center}
\caption{Ohmic dissipation rate, mass loss rate, and 
time scale for different relative angular velocities}
\end{figure} 

\begin{figure}[hbt]
\begin{center}
\epsscale{0.65}
\plotone{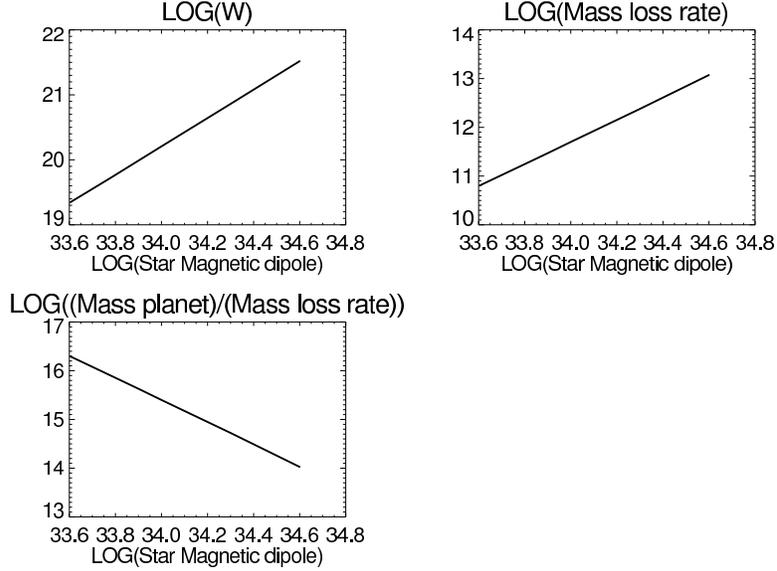}  
\label{Varying the star magnetic moment}
\end{center}
\caption{Ohmic dissipation rate, mass loss rate, 
and time scale for different stellar magnetic moment}
\end{figure}

\begin{figure}[hbt]
\begin{center}
\epsscale{0.65}
\plotone{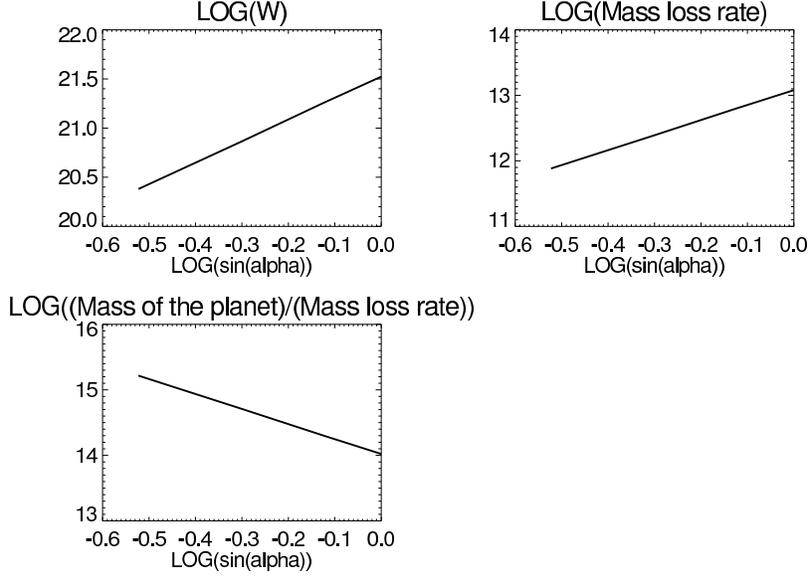}  
\end{center}
\caption{Ohmic dissipation rate, mass loss rate, 
and time scale for different tilt of the stellar magnetic dipole 
with regard to the stellar spin axis}
\label{Varying sin}
\end{figure}

\begin{figure}[hbt]
\begin{center}
\epsscale{0.65}
\plotone{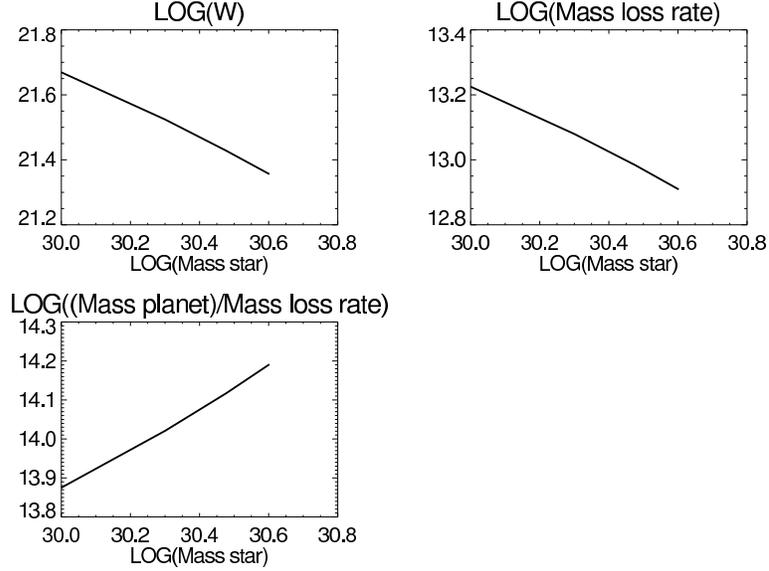}  
\label{Varying the stellar mass}
\end{center}
\caption{Ohmic dissipation rate, mass loss rate, 
and time scale for different stellar masses}
\end{figure}

\begin{figure}[hbt]
\begin{center}
\epsscale{0.65}
\plotone{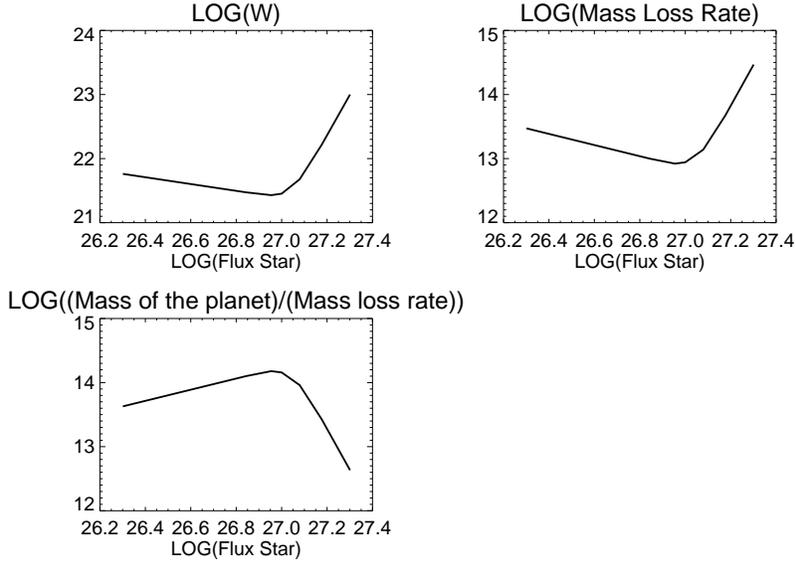}  
\label{Varying the stellar total flux}
\end{center}
\caption{Ohmic dissipation rate, mass loss rate, and time scale 
for different stellar total flux}
\end{figure}

From each group of plots, we obtain $\mathcal{P}$ and
$\stackrel{\centerdot}{M}$ as a function of the parameter that is
being varied (all the others are kept constant at the value of the
fiducial model given above).  These functions are given in the table
below (table \ref{functions}).

\begin{table}[htbp]
\begin{center}
\begin{tabular}{|c|c|}
\hline
Ohmic dissipation rate $\mathcal{P}$ and mass loss rate $\stackrel{\centerdot}{M}$ & Varying Parameter   \\
\hline
$\mathcal{P}_{1} = 3.3\times 10^{21}\ \left(\frac{M_{p}}{0.63\ M_{J}}\right)^{2.16}$ W&
$0.25\ M_{J} \leq M_{p} \leq 1.7\ M_{J}$  \\
$ \stackrel{\centerdot}{M}_{1} = 1.2\times 10^{13}\ \left(\frac{M_{p}}{0.63\ M_{J}}\right)^{2.4} {\rm kg \ s}^{-1}$  & \\
\hline
 $\mathcal{P}_{2} = 3.3\times 10^{21}\ \left(\frac{a}{0.04\ AU}\right)^{-4}$  W&
$0.015\ AU \leq a \leq\ 0.08\ AU$ \\
$ \stackrel{\centerdot}{M}_{2} = 1.2\times 10^{13}\ \left(\frac{a}{0.4\ AU}\right)^{-3.8} {\rm kg \ s}^{-1}$ & \\
\hline
$\mathcal{P}_{3} = 3.5\times 10^{21}\ \left(\frac{|\omega|}{10^{-5}}\right)-1.5\times 10^{20}$ W &
$7\times 10^{-6}\ s^{-1} \leq |\omega| \leq\ 7.3\times 10^{-5}\ s^{-1}$ \\
$ \stackrel{\centerdot}{M}_{3} = 1.4\times 10^{13}\ \left(\frac{|\omega|}{10^{-5}}\right)-1.3\times 10^{12} {\rm kg \ s}^{-1}$ & \\
\hline
$\mathcal{P}_{4} = 3.3\times 10^{21}\ \left(\frac{m}{4\times 10^{34}}\right)^{2.18}$ W & 
$6\times 10^{33}\ Am^{2} \leq m \leq\ 4\times 10^{34}\ Am^{2}$ \\
$ \stackrel{\centerdot}{M}_{4} = 1.2\times 10^{13}\ \left(\frac{m}{4\times 10^{34}}\right)^{2.3} {\rm kg \ s}^{-1}$ \\
\hline
$\mathcal{P}_{5}=3.3\times 10^{21} \sin^{2.17}(\alpha)$ W & 
$0.3 \leq \sin(\alpha) \leq 1$ \\
$\stackrel{\centerdot}{M}_{5} = 1.2\times 10^{13}\sin^{2.28}(\alpha) {\rm kg \ s}^{-1} $&  \\
\hline
$\mathcal{P}_{6} = 3.3\times 10^{21}\ \left(\frac{M_{\star}}{M_{\odot}}\right)^{-0.53}$ W & 
$0.5\ M_{\odot} \leq M_{\star} \leq\ 1.5\ M_{\odot}$ \\
$ \stackrel{\centerdot}{M}_{6} = 1.2\times 10^{13}\ \left(\frac{M_{\star}}{M_{\odot}}\right)^{-0.5} {\rm kg \ s}^{-1}$& \\
\hline
$\mathcal{P}_{7} = 3.3\times 10^{21}\ \left(\frac{L_{\star}}{1.5\ L{\odot}}\right)^{-0.5} $W & 
$0.5\ L_{\odot} \leq L_{\star} \leq\ 2.6\ L_{\odot}$ \\
$ \stackrel{\centerdot}{M}_{7} = 1.2\times 10^{13}\ \left(\frac{L_{\star}}{1.5\ L_{\odot}}\right)^{-0.8} {\rm kg \ s}^{-1}$& \\
$\mathcal{P}_{7} = 7.5\times 10^{19}\ \left(\frac{L_{\star}}{1.5\ L_{\odot}}\right)^{5.9} $W  & 
$2.6\ L_{\odot} \leq L_{\star} \leq\ 5\ L_{\odot}$ \\
$ \stackrel{\centerdot}{M}_{7} = 1.25\times 10^{11}\ \left(\frac{L_{\star}}{1.5\ L_{\odot}}\right)^{5.8} {\rm kg \ s}^{-1}$ & \\
\hline
\end{tabular}
\end{center}
\caption{Table giving $\mathcal{P}$ and $\stackrel{\centerdot}{M}$ as  
a function of the parameter that is being varied.}
\label{functions}
\end{table}

\subsection{Model parameter dependence}
\begin{enumerate}
\item \textbf{Mass of the planet $\mathbf{M_{p}}$.}  The total ohmic
dissipation is a volumic integral over the entire region where
dissipation occurs.  Therefore, one might expect $\mathcal{P}$ to be
proportional to the volume.  Since the planet fills its Roche lobe,
the volume is detemined by the mass (cf. eq. \ref{definition hill's
radius}).  However, $M_{p}$ also gives a constraint on the volumic
mass at the center of the planet (cf. equation \ref{3 unknowns}),
which makes $\mathcal{P}$ mostly proportional to $M^{2}$. There is
also a minor correction due to $\mathcal{P}$'s weak dependence on
$\eta$ which depends on $M_{p}$ through the calculation of the
internal parameters $T$, $P$, $\rho$ (eqs. \ref{P isothermal model}
and \ref{P polytropic model}).

\item \textbf{Semi major axis $\mathbf{a}$.}  In the model we adopted in \S4,
or more generally, in a model that would not take into account the
planet's internal adjustment to the ohmic dissipation (especially in a
model in which the radius of the planet is independent of the
semi-major axis), the ohmic dissipation rate would be related to the
semi-major axis according to the following law: $\mathcal{P} \propto
B^{2} \propto a^{-6}$.  In the self-consistent model we adopted here
and in the limit where the planet fills its Roche lobe
($R_{p}=R_{H}$), the radius of the planet varies with the semi-major
axis. For example, when a planet moves closer to its host star, its
Roche radius decreases linearly with the radius
(cf. eq. \ref{definition hill's radius}) and, therefore, $\mathcal{P}$
increases less quickly than if the planet kept the same radius.
Again, $\mathcal{P}$ also has a weak dependence on $\eta$ which is
depends on the semi-major axis through the planet's surface
temperature and through the dependence of $r_{2}$ on $a$.  From these
arguments, we thus expect $\mathcal{P}$ the exponent in
$\mathcal{P}_{2}$ to be greater than -6 and less than -3.

\item \textbf{Relative angular velocity $\mathbf{\omega}$.}  In equation
(\ref{dependence of P)}, the multiplicative constant in front
of the volumic integral comes comes from the induction by the time
dependent stellar field (cf. eq. (\ref{rhs=lhs}) and eq. (\ref{W
provisoire})) and gives to $\mathcal{P}$ a dependence on
$\omega^{2}$. However, in addition, $\omega$ intervenes inside the
volumic integral in eq. \ref{dependence of P} through the
dependence of $G_{l}(r)$ on the $\delta^{-1}$, with
$\delta=\sqrt{\frac{\eta}{\omega}}$, skin depth. Therefore, the
volumic integral is proportional to $\omega^{-1}$ and 
$\mathcal{P}$ is proportional to $\omega$.

\item \textbf{Magnetic dipole $\mathbf{m}$ and tilt of the magnetic dipole
$\alpha$.}  Without any adjustment of the planet's interior to the
ohmic dissipation, we would expect $\mathcal{P}_{4}$ and
$\mathcal{P}_{5}$ to vary respectively in $m^{2}$ and
$\sin^{2}(\alpha)$. The fact that both exponents that have been
computed numerically are slightly larger than 2 means, in the
self-consistent model we adopted here, that the adjustment of the
planet's interior tends to have a small positive retroaction on the
amount of enegy that is deposited inside the planet through ohmic
dissipation.

\item \textbf{Mass of the star $\mathbf{M_{\star}}$.}  The mass of the star
intervenes in the computation of the Roche radius ($R_{H} \propto
M_{\star}^{-1/3}$).  $\mathcal{P}$ being a volumic integral, one
would, therefore, expect it to vary proportionally to
$M_{\star}^{-1}$.  However, $M_{\star}$ also intervenes in the
computation of $r_{2}$ ($r_{2}$ is the sonic point, or the largest
solution of $(r^{2}-rR_{H}+\frac{2c_{s}^{2}a^{3}}{LGM_{\star}})=0$,
see. eq \ref{l}), which brings a correction to the dependence of
$\mathcal{P}$ on $M_{s}$. Indeed, $r_{2}$ is used to compute
$P(r_{2})$ and $\rho(r_{2})$ (cf. eq. \ref{P(r2)}) which are the
boundary conditions we adopted to calculate the pressure and volumic
mass in the isothermal region (cf. eq. \ref{P in isothermal
region}). As a side note, this dependence of $\mathcal{P}$ on $r_{2}$
could also affect the dependence of $\mathcal{P}$ on $a$ and
$M_{\star}$.)

\item \textbf{Stellar total Luminosity $\mathbf{L_{\star}}$.}  From figure 9,
one can see that $\L_{\star}=10^{27}\ W$ correponds to a minimum for
$\mathcal{P}$ and that $\mathcal{P}$ varies slowly for $\L_{\star}\leq
10^{27}\ W$ and much faster for $\L_{\star} \geq 10^{27}\ W$.  In the
model we adopted here, the stellar total luminosity fixes the planet's
equilibrium surface temperature, which is also the temperature of the
isothermal region (for $\L_{\star}=10^{27}\ W$, $T_{p}\simeq 1767 K.$).
It in turns determines the temperature profile inside the planet (the
surface temperature is used as a boundary condition) and influences
the internal structure and magnetic diffusivity profile inside the
planet $\eta(r)$.  $T_{p}$ varies proportionally to $L_{\star}^{1/4}$
(for constant semi-major axis) and, therefore, $\eta$ is roughly
proportional to $exp(\dfrac{78909}{L_{\star}^{1/4}})$. \\

We ploted the integrand of the ohmic dissipation ($< \mathcal{P}_{volumic} > r^{2}$ 
in eq. \ref{energy_integrand_expression} ) as well as the
magnetic diffusivity (see figures 10 and 11) for
$L_{\star}=2\times 10^{26}$ W, $7\times 10^{26}$ W, $10^{27}$ W,
$1.2\times 10^{27}$ W, $1.5\times 10^{27}$ W, 
and $2\times 10^{27}$ W respectively.  One can notice two parts
corresponding to the isothermal and the polytropic region.  In the
isothermal region, the energy integrand decreases slowly from the
surface to the center of the planet.  The transition between the two
regions corresponds to a sharp increase in conductivity and,
therefore, a quick increase in the ohmic dissipation. This accounts
for the sharp increase in the integrand (around $1.5\times 10^8$
meters), and most of the remaining magnetic energy is dissipated in
this region.  Moreover, an increase in the stellar luminosity results
in a decrease in the magnetic diffusivity as well as a deeper
penetration and a deeper transition between the isothermal and the
polytropic region.\\

When one increases $L_{\star}$ starting from low values
($L_{\star}=2\times 10^{26}$W), the energy integrand also
increases in the isothermal region. Nevertheless, the extent and amplitude of the sharp
increase at the transition between the isothermal and the polytropic
region are also reduced.  Therefore, these two effects compensate each
other, and for low stellar luminosity (e.g. $L_{\star}$ between
$2\times 10^{26}$W and $10^{27}$W, the total ohmic dissipation in the
planet increases slowly with the stellar luminosity). \\

On the other hand, for higher values of the stellar luminosity, the
penetration depth as well as the transition depth saturates (when the coupling term in eq. (\ref{first equation on G})
$\dfrac{\omega}{\eta(r)}$ between the real and imaginary parts of $G_{l}(r)$ 
becomes comparable to the other term $(\dfrac{l(l+1)}{r^{2}})$).  One can
see that an increase in the stellar luminosity results only in changes
in the energy integrand that would increase the total ohmic
dissipation. This result in a much sharper increase of the ohmic
dissipation with the stellar luminosity.

\begin{figure}[hbt]
\begin{center}
\epsscale{0.75}
\plotone{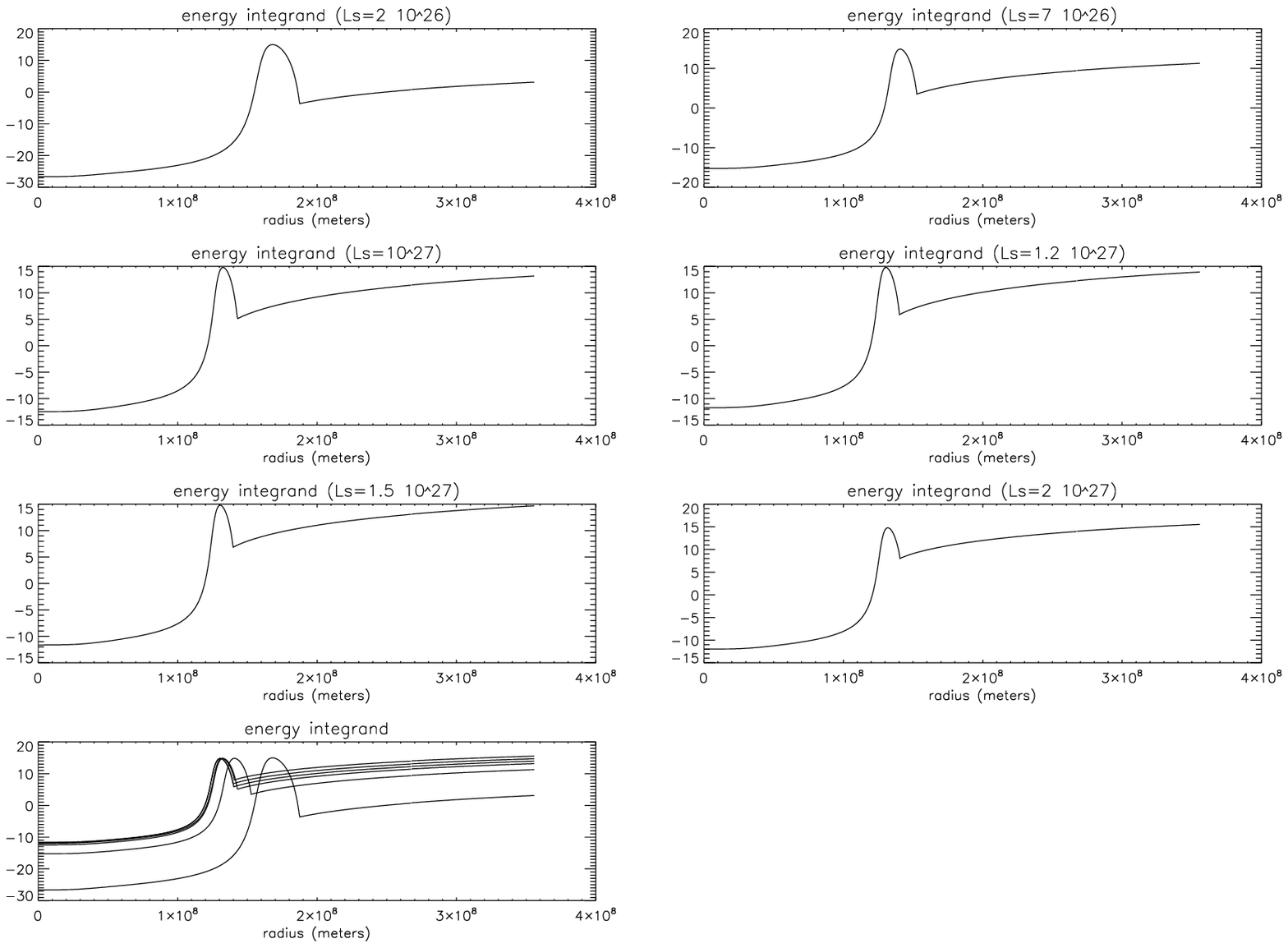}  
\label{energy_integrand_plot}
\end{center}
\caption{Integrand of the ohmic dissipation (log scale) for different values of the stellar luminosity}
\end{figure}

\begin{figure}[hbt]
\begin{center}
\epsscale{0.75}
\plotone{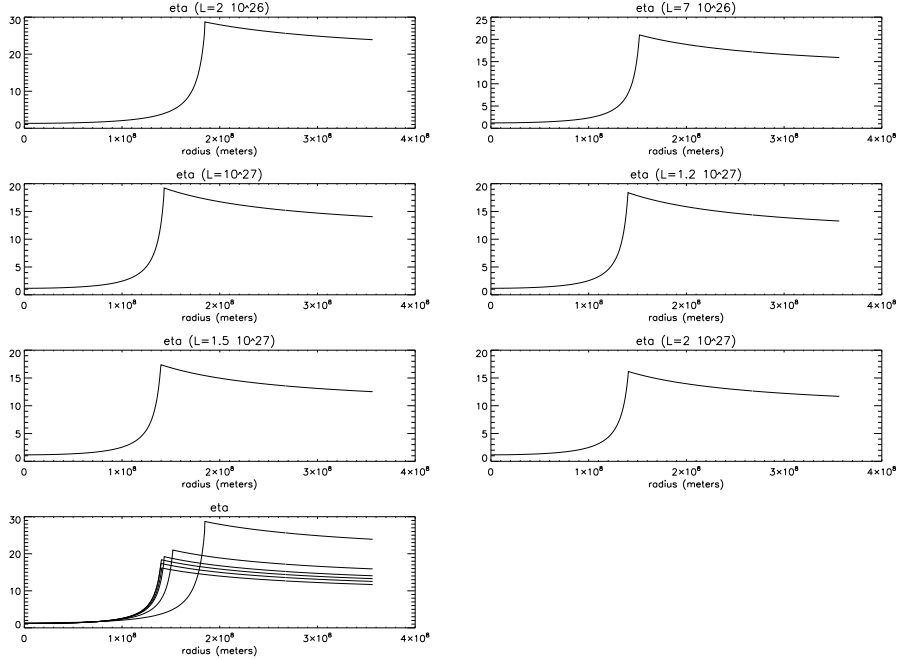}  
\label{plot_of_eta}
\end{center}
\caption{Magnetic diffusivity (log scale) for different values of the stellar luminosity}
\end{figure}

\item \textbf{Attempts of generalized function.}
We consider the generalized expression of the ohmic dissipation rate 
and mass loss rate, in the case 
where the variables are separable: 
\begin{eqnarray}
\mathcal{P}_{gen}=3.3\times 10^{21} W
\left(\frac{M_{p}}{0.63\ M_{J}}\right)^{2.16} 
\left(\frac{a}{0.04\ AU}\right)^{-4}
\left[
\left(\frac{|\omega|}{10^{-5} {\rm s}^{-1}}
-0.045\right)\right]\nonumber \\
\left(\frac{m}{4\times 10^{34} {\rm A \ m}^2 }\right)^{2.18}
\left(\sin\ \alpha \right)^{2.17}
\left(\frac{M_{\star}}{M_{\odot}}\right)^{-0.53}
\left(\frac{L_{\star}}{1.5\ L{\odot}}\right)^{-0.5} 
\label{generalized P}
\end{eqnarray}

\begin{eqnarray}
\mathcal{\stackrel{\centerdot}{M}}_{gen}=1.2\times 10^{13} {\rm kg \
s}^{-1} \left(\frac{M_{p}}{0.63\ M_{J}}\right)^{2.4}
\left(\frac{a}{0.04\ AU}\right)^{-3.8}
\left[\left(\frac{|\omega|}{10^{-5} {\rm s}^{-1}} -0.1
\right)\right]\nonumber \\ \left(\frac{m}{4\times 10^{34} {\rm A \
m}^2 }\right)^{2.3} \left(\sin\ \alpha \right)^{2.28}
\left(\frac{M_{\star}}{M_{\odot}}\right)^{-0.5}
\left(\frac{L_{\star}}{1.5\ L{\odot}}\right)^{-0.8}. 
\label{generalized Mdot}
\end{eqnarray}

The previous formulas have been written for the first interval of 
$\mathcal{P}_{7}$ in table 2, but one can write the corresponding formulas for 
the second interval of by using the corresponding expression of $\mathcal{P}_{7}$. 
We compute, using the iterative procedure described above (\S 7.2),
the ohmic dissipation $\mathcal{P}$ and mass loss rate
$\mathcal{\stackrel{\centerdot}{M}}$ for sets of parameters in which
more than 2 parameters are different from the fiducial parameters.  We
then compare these values with the value of $\mathcal{P}_{gen}$ and
$\mathcal{\stackrel{\centerdot}{M}}_{gen}$ for these sets of
parameters. This test enables us to determine if the hypothesis of
separation of variable is accurate or not.  The first set of
parameters that we consider is $M_{p}=1.5 M_{J}$ and $ a=0.03 AU$, all
the other parameters being kept equal to the fiducial parameters. We
get: $\mathcal{P}=4.2\times 10^{22}$W and
$\mathcal{\stackrel{\centerdot}{M}}=1.5\times 10^{14} {\rm kg \
s}^{-1}$.  However, using the formula of the ohmic dissipation rate
and mass loss rate with the approximation of separation of variables,
we get: $\mathcal{P}_{gen}=7\times 10^{22}$W and
$\mathcal{\stackrel{\centerdot}{M}}=2.9\times 10^{14}{\rm kg \ s}^{-1}$.

We now consider a set of parameters in which all parameters are taken
differnet from the fiducial value: $M_{p}=1.5M_{j},\ a=0.03AU,\
\omega=2.9\times 10^{-5}s^{-1},\ m=2\times 10^{34}\ Am^{2},\ M_{\star}=1.5
M_{\odot}, \L_{\star}=0.8 \L_{\odot}$ (this value of $\omega$
corresponds, for example to a system in which the planet is at
keplerian augular velocity and the star has a period of 4 days).  We
get $\mathcal{P}=1.7\times 10^{22}$W and
$\mathcal{\stackrel{\centerdot}{M}}=7\times 10^{13} {\rm kg \
s}^{-1}$, and using the formula of the ohmic dissipation rate and mass
loss rate with the approximation of separation of variables, we get
$\mathcal{P}_{gen}=3.5\times 10^{22}$W and
$\mathcal{\stackrel{\centerdot}{M}}=1.6\times 10^{14} {\rm kg \
s}^{-1}$. \\

From these two tests, we deduce that the approximation of separation
of variables gives a reasonable order of magnitude, but nevertheless,
does not seem to be accurate. This results means that the exponents in
the functions given in table 2 can themselves be a function of the
parameters, and the generalized functions that describe the ohmic
dissipation and mass loss rate as a function of all parameters can be
fairly complicated.  Nevertheless, even if a generalized formula is
still unknown, for a given set of parameters, one could still compute
the mass loss rate and ohmic dissipation using the procedure described
in \S\S 3-7.
\end{enumerate}

\subsection{Mass loss and migration stalls}
From equation (\ref{generalized Mdot}), we find
\begin{eqnarray}
\tau_m= \frac{M_p}{\mathcal{\stackrel{\centerdot}{M}}_{gen}}
\simeq 3 {\rm Myr} \left(\frac{0.63\ M_{J}}{M_{p}}\right)^{2.4}
\left(\frac{a}{0.04\ AU}\right)^{3.8}
\left[\left(\frac{|\omega|}{10^{-5} {\rm s}^{-1}} -0.1 \right)\right]^{-1}
\left(\frac{m}{4\times 10^{34} {\rm A m}^2}\right)^{-2.3} \nonumber \\
\left(\sin\ \alpha \right)^{-2.28}   
\left(\frac{M_{\star}}{M_{\odot}}\right)^{0.5}
\left(\frac{L_{\star}}{1.5\ L{\odot}}\right)^{0.8} \label{taum}
\end{eqnarray}

The same mass loss provides angular momentum to the planet.  We
neglect any variations in excentricity and assume that all the mass is
accreted into the star.  Using (GBL eq. 96), we link the mass loss
rate to a rate of change of semi-major axis,
\begin{equation}
\frac{\stackrel{\centerdot}{a}}{a}=-2
\frac{\stackrel{\centerdot}{M}} {Mp} 
\label{rate of semi-major axis change}
\end{equation}
Thus, $\tau_a \equiv \vert {\dot a} \vert /a = - 2 \tau_m$.

These relations indicate that, within $\sim 0.04$ AU, the Ohmic
dissipation within the planet may indeed generate sufficient energy to
inflat their radius beyond their Roche lobe.  The resulting mass
transfer not only reduces the planets' mass but also stall their
orbital migration.  The impact of this process on the mass-period 
distribution of gas giants will be discussed in paper III.

\section{Summary}
In this paper, we applied a model described by Campbell (in the
context of interacting binary stars) to the situation of a planet in a
protoplanetary disk interacting with the stellar periodic magnetic
field.  In \S 3, we showed that with a well determined electrical
conductivity profile inside the planet as well as the characteristic
parameters of the system (such as the stellar magnetic field strength
and angular velocity spin, planet radius and semi-major axis), one can
compute the total ohmic dissipation rate $\mathcal{P}(t)$ inside the
planet and its average value over one synodic period. This dissipation
rate gives a good estimate of the strength of the Lorrentz torque
exerted on the planet due to the interaction between the stellar
magnetic field and the induced current inside the planet.  When the
planet is outside corrotation, this torque will provide angular
momentum to the planet from the star and slow down the planet's
migration.  In \S4, we computed $\mathcal{P}$ for one specific set of
parameters ($R_{p}=0.63\ R_{J}$, $a=0.04$ AU), and also showed that
the conductivity profile (all the other parameters being kept
constant) had some influence on the location of maximum dissipation,
but fairly little influence on the total dissipate rate $\mathcal{P}$.
We noted that this value of $\mathcal{P}$ seemed too low to directly
provide an adequate rate of angular momentum transfert to the planet
to stop its migration toward the host star. However, this energy input
can inflat the planet's enveloppe and trigger mass loss
$\stackrel{\centerdot}{M}$ through Roche lobe overflow.  The mass that
overflows toward the central star provides angular momentum to the
planet (GBL). In order to estimate this mass loss rate, we first
linked the ohmic dissipation rate to the mass loss rate (\S5).  Then
we used an isotermal-polytropic model to describe the adjustment of
the planet's interior to the heat deposited through ohmic dissipation
(\S6).  Finally, we computed $\mathcal{P}$ and
$\stackrel{\centerdot}{M}$ at equilibrium for several set of
parameters (\S7). A detailed calculation on the orbital evolution of
the planet due to this process will be presented in paper III.

\acknowledgments

We thank Drs. P. Garaud, J.E. Pringle and F. Rasio for constructive
discussions. This work is supported by NASA (NAGS5-11779, NNG04G-191G,
NNG06-GH45G, NNX07AL13G, HST-AR-11267), JPL (1270927), and NSF(AST-0507424).

\appendix
\section{Perfect conductor moving relative to a magnetic field}
Let's consider the flux $\Phi$ of the magnetic field $\textbf{B}$ accross a surface $S(t)$ that 
changes with time or moves in space. One can show that 
\begin{eqnarray}
\frac{d\Phi}{dt}\stackrel{def}{=}\frac{1}{dt}\left[\int_a \textbf{B}(\textbf{r},t+dt)\, d\mathbf{S} 
-\int_b \mathbf{B}(\textbf{r},t)\, d\mathbf{S} \right]
=\int_a \left[\frac{\partial\textbf{B}}{\partial t}-\nabla\wedge(\upsilon\wedge
\textbf{B})\right]\, d\mathbf{S}
\end{eqnarray}
with $a=S(t+dt)$, and $b=S(t)$. Therefore, using the mhd induction equation 
\begin{equation}
\frac{\partial\textbf{B}}{\partial t}=
\nabla\wedge(\upsilon\wedge
\textbf{B})
-\nabla\wedge(\dfrac{1}{\mu_{0}\sigma}\nabla\wedge\textbf{B}),
\end{equation} 
we get: 
\begin{equation}
\frac{d\Phi}{dt}=-\frac{1}{\sigma}\int_c \mathbf{J}(\textbf{r},t)\, d\mathbf{l}
\end{equation} 
which tends to zero when the electric conductivity is large (the previous integral is 
a closed integral along a close curve). Therefore, the magnetic field's flux will be constant if $\sigma$ 
is large enough that the second term in the right hand side 
is negligible. In such a case, the field lines will move with the body and appear to be 'frozen.'

\section{Set of linear equations}
\label{set of linear equations}
We give below the linear set of equation that we solved for  
($\mu_{1}^{1}$, $\mu_{1}^{-1}$, $\nu_{1}^{1}$,
$\nu_{1}^{-1}$, $\alpha_{1}$, $\alpha_{2}$,$\alpha_{3}$,$\alpha_{4}$),
($\mu_{2}^{0}$, $\nu_{2}^{0}$, $\beta_{1}$,$\beta_{2}$), and
($\mu_{2}^{2}$, $\mu_{2}^{-2}$, $\nu_{2}^{2}$, $\nu_{2}^{-2}$,
$\gamma_{1}$, $\gamma_{2}$, $\gamma_{3}$, $\gamma_{4}$). 
The values of $G_{l}(r)$ considered are for $r=R_{p}$ the radius of the planet. 

\begin{eqnarray}
\left\{
\begin{array}{l}
(\mu_{1}^{1}-\mu_{1}^{-1})Re(G_{1})+(-\nu_{1}^{1}+\nu_{1}^{-1})Im(G_{1}) -\alpha_{2}\frac{1}{R_{p}}\sqrt{\frac{8\pi}{3}}=0\\
(-\nu_{1}^{1}-\nu_{1}^{-1})Re(G_{1})+(-\mu_{1}^{1}-\mu_{1}^{-1})Im(G_{1}) -\alpha_{4}\frac{1}{R_{p}}\sqrt{\frac{8\pi}{3}}=\frac{m\ \sin\alpha}{8\pi\ d^{3}}R_{p}^{2}\sqrt{\frac{8\pi}{3}}\\
(-\nu_{1}^{1}+\nu_{1}^{-1})Re(G_{1})+(-\mu_{1}^{1}+\mu_{1}^{-1})Im(G_{1}) -\alpha_{1}\frac{1}{R_{p}}\sqrt{\frac{8\pi}{3}}=2\frac{m\ \sin\alpha}{8\pi\ d^{3}}R_{p}^{2}\sqrt{\frac{8\pi}{3}}\\
(-\mu_{1}^{1}-\mu_{1}^{-1})Re(G_{1})+(\nu_{1}^{1}+\nu_{1}^{-1})Im(G_{1}) -\alpha_{3}\frac{1}{R_{p}}\sqrt{\frac{8\pi}{3}}=0 \\
(\mu_{1}^{1}-\mu_{1}^{-1})Re(\stackrel{\centerdot}{G}_{1})+(-\nu_{1}^{1}+\nu_{1}^{-1})Im(\stackrel{\centerdot}{G}_{1})+ \alpha_{2}\frac{1}{R^{2}_{p}}\sqrt{\frac{8\pi}{3}}=\frac{m\ \sin\alpha}{8\pi\ d^{3}}2R_{p}\sqrt{\frac{8\pi}{3}}\\
(-\nu_{1}^{1}-\nu_{1}^{-1})Re(\stackrel{\centerdot}{G}_{1})+(-\mu_{1}^{1}-\mu_{1}^{-1})Im(\stackrel{\centerdot}{G}_{1})+ \alpha_{4}\frac{1}{R^{2}_{p}}\sqrt{\frac{8\pi}{3}}=2\frac{m\ \sin\alpha}{8\pi\ d^{3}}2R_{p}\sqrt{\frac{8\pi}{3}}\\
(-\nu_{1}^{1}+\nu_{1}^{-1})Re(\stackrel{\centerdot}{G}_{1})+(-\mu_{1}^{1}+\mu_{1}^{-1})Im(\stackrel{\centerdot}{G}_{1})+ \alpha_{1}\frac{1}{R^{2}_{p}}\sqrt{\frac{8\pi}{3}}=0 \\
(-\mu_{1}^{1}-\mu_{1}^{-1})Re(\stackrel{\centerdot}{G}_{1})+(\nu_{1}^{1}+\nu_{1}^{-1})Im(\stackrel{\centerdot}{G}_{1})+ \alpha_{3}\frac{1}{R^{2}_{p}}\sqrt{\frac{8\pi}{3}}=0\\
\end{array}
\right.
\end{eqnarray}
\begin{eqnarray}
\left\{
\begin{array}{l}
(\mu_{2}^{2}+\mu_{2}^{-2})Re(G_{2}(R_{p}))+(-\nu_{2}^{2}-\nu_{2}^{-2})Im(G_{2}(R_{p})) -\gamma_{2}\frac{1}{R_{p}^{2}}12\sqrt{\frac{2\pi}{15}}=0\\
(-\nu_{2}^{2}+\nu_{2}^{-2})Re(G_{2}(R_{p}))+(-\mu_{2}^{2}+\mu_{2}^{-2})Im(G_{2}(R_{p})) -\gamma_{4}\frac{1}{R_{p}^{2}}12\sqrt{\frac{2\pi}{15}}=-\frac{1}{3}\dfrac{m\ \sin\alpha}{8\pi\ d^{4}}R_{p}^{3}12\sqrt{\dfrac{2\pi}{15}}\\
(-\nu_{2}^{2}-\nu_{2}^{-2})Re(G_{2}(R_{p}))+(-\mu_{2}^{2}-\mu_{2}^{-2})Im(G_{2}(R_{p})) -\gamma_{1}\frac{1}{R_{p}^{2}}12\sqrt{\frac{2\pi}{15}}=-\frac{1}{2}\dfrac{m\ \sin\alpha}{8\pi\ d^{4}}R_{p}^{3}12\sqrt{\dfrac{2\pi}{15}}\\
(-\mu_{2}^{2}+\mu_{2}^{-2})Re(G_{2}(R_{p}))+(\nu_{2}^{2}-\nu_{2}^{-2})Im(G_{2}(R_{p})) -\gamma_{3}\frac{1}{R_{p}^{2}}12\sqrt{\frac{2\pi}{15}}=0\\
(\mu_{2}^{2}+\mu_{2}^{-2})Re(\stackrel{\centerdot}{G}_{2}(R_{p}))+(-\nu_{2}^{2}-\nu_{2}^{-2})Im(\stackrel{\centerdot}{G}_{2}(R_{p})) +2\gamma_{2}\frac{1}{R_{p}^{3}}12\sqrt{\frac{2\pi}{15}}=0\\
(-\nu_{2}^{2}+\nu_{2}^{-2})Re(\stackrel{\centerdot}{G}_{2}(R_{p}))+(-\mu_{2}^{2}+\mu_{2}^{-2})Im(\stackrel{\centerdot}{G}_{2}(R_{p})) +2\gamma_{4}\frac{1}{R_{p}^{3}}12\sqrt{\frac{2\pi}{15}}=-\dfrac{m\ \sin\alpha}{8\pi\ d^{4}}R_{p}^{2}12\sqrt{\dfrac{2\pi}{15}}\\
(-\nu_{2}^{2}-\nu_{2}^{-2})Re(\stackrel{\centerdot}{G}_{2}(R_{p}))+(-\mu_{2}^{2}-\mu_{2}^{-2})Im(\stackrel{\centerdot}{G}_{2}(R_{p})) +2\gamma_{1}\frac{1}{R_{p}^{3}}12\sqrt{\frac{2\pi}{15}}=-\frac{3}{2}\dfrac{m\ \sin\alpha}{8\pi\ d^{4}}R_{p}^{2}12\sqrt{\dfrac{2\pi}{15}}\\
(-\mu_{2}^{2}+\mu_{2}^{-2})Re(\stackrel{\centerdot}{G}_{2}(R_{p}))+(\nu_{2}^{2}-\nu_{2}^{-2})Im(\stackrel{\centerdot}{G}_{2}(R_{p})) +2\gamma_{3}\frac{1}{R_{p}^{3}}12\sqrt{\frac{2\pi}{15}}=0
\end{array}
\right.
\end{eqnarray}
\begin{eqnarray}
\left\{
\begin{array}{l}
\mu_{2}^{0}Re(G_{2}(R_{p}))-\nu_{2}^{0}Im(G_{2}(R_{p})) -\beta_{2}\frac{1}{R_{P}^{2}}\sqrt{\frac{4\pi}{5}}=0\\
-\nu_{2}^{0}Re(G_{2}(R_{p}))-\mu_{2}^{0}Im(G_{2}(R_{p})) -\beta_{1}\frac{1}{R_{P}^{2}}\sqrt{\frac{4\pi}{5}}=\sqrt{\dfrac{4\pi}{5}}\dfrac{m\ \sin\alpha}{8\pi\ d^{4}}R_{p}^{3} \\
\mu_{2}^{0}Re(\stackrel{\centerdot}{G}_{2}(R_{p}))-\nu_{2}^{0}Im(\stackrel{\centerdot}{G}_{2}(R_{p})) +\beta_{2}\frac{2}{R_{P}^{3}}\sqrt{\frac{4\pi}{5}}=0\\
-\nu_{2}^{0}Re(\stackrel{\centerdot}{G}_{2}(R_{p}))-\mu_{2}^{0}Im(\stackrel{\centerdot}{G}_{2}(R_{p})) +\beta_{1}\frac{2}{R_{P}^{3}}\sqrt{\frac{4\pi}{5}}=\sqrt{\dfrac{4\pi}{5}}\dfrac{m\ \sin\alpha}{8\pi\ d^{4}}3R_{p}^{2}
\end{array}
\right.
\end{eqnarray}

\section{Coefficients intervening in the expression of the ohmic dissipation rate}
\begin{eqnarray}
\left\{
\begin{array}{l}
A_{11}(r)\stackrel{def}{=}(\nu_{1}^{1}-\nu_{1}^{-1})Re(G_{1}(r))+(\mu_{1}^{1}-\mu_{1}^{-1})Im(G_{1}(r)) \\
A_{12}(r)\stackrel{def}{=}(\mu_{1}^{1}+\mu_{1}^{-1})Re(G_{1}(r))-(\nu_{1}^{1}+\nu_{1}^{-1})Im(G_{1}(r)) \\
A_{13}(r)\stackrel{def}{=}(\mu_{1}^{1}-\mu_{1}^{-1})Re(G_{1}(r))-(\nu_{1}^{1}-\nu_{1}^{-1})Im(G_{1}(r)) \\
A_{14}(r)\stackrel{def}{=}-(\nu_{1}^{1}+\nu_{1}^{-1})Re(G_{1}(r))-(\mu_{1}^{1}+\mu_{1}^{-1})Im(G_{1}(r)) \\
A_{15}(r)\stackrel{def}{=}\nu_{2}^{0}Re(G_{2}(r))+\mu_{2}^{0}Im(G_{2}(r)) \\
A_{16}(r)\stackrel{def}{=}\mu_{2}^{0}Re(G_{2}(r))-\nu_{2}^{0}Im(G_{2}(r)) \\
A_{17}(r)\stackrel{def}{=}(\nu_{2}^{2}+\nu_{2}^{-2})Re(G_{2}(r))+(\mu_{2}^{2}+\mu_{2}^{-2})Im(G_{2}(r)) \\
A_{18}(r)\stackrel{def}{=}(\mu_{2}^{2}-\mu_{2}^{-2})Re(G_{2}(r))+(-\nu_{2}^{2}+\nu_{2}^{-2})Im(G_{2}(r)) \\
A_{19}(r)\stackrel{def}{=}(\mu_{2}^{2}+\mu_{2}^{-2})Re(G_{2}(r))-(\nu_{2}^{2}+\nu_{2}^{-2})Im(G_{2}(r)) \\
A_{20}(r)\stackrel{def}{=}(-\nu_{2}^{2}+\nu_{2}^{-2})Re(G_{2}(r))-(\mu_{2}^{2}-\mu_{2}^{-2})Im(G_{2}(r)) 
\end{array}
\right.
\end{eqnarray}

\section{Conductivity and resistance of short-period extrasolar planets}

We now calculate the conductivity and resistance (the reciprocal of
conductivity) of short-period extrasolar planets.  The conductivity of
the planet is determined by the density of charged particles.  
We consider separately the contributiond from the collisional 
ionization within the planet's interior and from the photoionization
near its surface.  

\subsection{Ionization of the planet's interior}

The cores and the inner envelopes of Jovian planets are mostly
ionized, they are shielded by cool, mostly neutral gaseous envelopes,
where the ionization is dominated by elements with low ionization
potentials, such as sodium and potassium.

The planetary model used in our calculation of ionization fraction is
model A3 presented by Bodenheimer et al. (2001). It is a spherically
symmetric model for the short-period planet around HD 209458. The
following parameters are assumed: the planetary mass is 0.63 Jupiter
masses ($M_J$); the equlibrium temperature at the surface of the
planet due to irradiation from the star is $T_s = 1360$~K; there is a
solid core with a density $\rho_c = 5.5\ 10^{3} \ \rm{kg \ m^{-3}}$ and a mass
0.139 $M_J$ ( $ = 44 M_{\oplus} = 0.22 M_p $ ) in the center. An
energy source, uniformly distributed through the gaseous part of the
planet, with an energy input rate $\dot{E}_d = 8.5 \times 10^{19} \
\rm{J \ s^{-1}}$, is also imposed to take into account the effect
of tidal dissipation of energy. Those model parameters result in an
asymptotic radius of $1.41 R_J$ at $t = 4.5$~Gyr, which is consistent
with the photometric occultation observations of HD 209458 (Henry et
al. 2000b; Charbonneau et al. 2000)

The cores and the inner envelope of Jovian planets are mostly ionized
by the pressure ionization effect, where the Saha equation breaks
down.  We have to resort to various equations of state (EOS) tables,
including the equation of state tables for hydrogen and helium by
Saumon et al. (1995).  These tables are calculated using the free
energy minimization methods, with a careful study of the nonideal
interactions.  They cover temperatures in the range $2.10 < \log
T({\rm K}) < 7.06$ and pressure in the range $5 < \log P ({\rm N \
m^{-2}}) < 20$.  The calculations based on which these tables were
constructed also include the treatments of partial dissociation and
ionization caused by both pressure and temperature effects. Given the
internal structure data of $\rho$, $T$ and $P$ for model A3, we use
those EOS tables to calculate electron number density distribution for
the inner part of the planet.  In this approximation, we bear in mind
that the ionization fraction calculated from the free-energy
minimization are of limited accuracy.  Moreover, in the interpolation
regions of both the H and He equation of state, the data have very
little physical basis but are reasonably well behaved by construction.

In the envelope, hydrogen and helium are mostly neutral, and free
electrons are exclusively provided by thermal ionization of elements
with low ionization potentials. Among them, potassium and sodium have
highest concentrations with a relative abundance $\log (N_K / N_H)
\simeq -6.88$ and $\log (N_{Na} / N_H) \simeq -5.67$ for the solar
composition.  The lowest ionization potentials for these two elements
are $\chi_{K} = 4.339 $~eV and $\chi_{Na} = 5.138 $~eV. We identify
these two elements to be the sources of most of the free electrons in
the planetary envelope.

We solve the Saha equations for ionization of Na \& K jointly
(cf. Allen 1955)
\begin{equation}
\frac{N^1_{Na}}{N^0_{Na}}\ 10^{6}\times N_e = - \chi_{Na} \Theta - {\frac{3}{2}} \log
\Theta + 20.9388
\end{equation}
and
\begin{equation}
\frac{N^1_{K}}{N^0_{K}}\ 10^{6}\times N_e = - \chi_{K} \Theta - {\frac{3}{2}} \log
\Theta + 20.9388
\end{equation}
where $\Theta = 5040 K / T$, $N^1_{Na}$ and $N^0_{Na}$ are,
respectively, the singly ionized and neutral number density of sodium (SI units),
$N^1_{K}$ and $N^0_{K}$ are, respectively, the singly ionized and
neutral number density of potassium, and $N_e$ is the electron
density.

\subsection{Conductivity and resistivity}

Using the electron number density profile, we then calculate the
conductivity using the formulas given by Fejer (1965). Three kinds of
conductivity are of interests here. The conductivity $\sigma_0$, which
determines the current parallel to the magnetic lines of force, and
which would exist for all direction in the absence of the magnetic
field, is given by
\begin{equation}
\sigma_0 = \frac{N_e e }{B} \left(\frac{\omega_{i}}{\nu_{i}} 
-\frac{\omega_{e}}{\nu_{e}} \right)
\cong \frac{N_e e^2}{m_e \nu_{e}} 
\label{eq:sigma0}
\end{equation}
where $e$ is the electron charge, $\omega_{e} = - \frac{e B}{m_{e}}$
and $\omega_{i} = \frac{e B}{m_{i}}$ are the gyrofrequencies of
electron and ion, respectively, while $\nu_{e}$ and $\nu_{i}$ are the
collisional frequencies associated with the momentum transfer of
electrons and ions. 

The Pedersen conductivity, which determines current parallel to the 
electric field, is given by  
\begin{equation}
\sigma_p = \frac{N_e e }{B} \left( \frac{\nu_{i} \omega_{i}}{\nu_{i}^{2}+
\omega_{i}^{2}}  - \frac{\nu_{e} \omega_{e}}{\nu_{e}^{2}+
\omega_{e}^{2}} \right)
\cong \frac{\sigma_0 }{1 + (\omega_e / \nu_e)^2}.
\label{eq:sigmap}
\end{equation}
The Hall conductivity, which determines the current perpendicular to both the
electric and magnetic fields, is given by
\begin{equation}
\sigma_h = \frac{N_e e }{B} \left( \frac{\omega_{e}^{2}}{\nu_{e}^{2}+
\omega_{e}^{2}} - \frac{\omega_{i}^{2}}{\nu_{i}^{2}+
\omega_{i}^{2}} \right) 
\cong \frac{\sigma_0 (\omega_e / \nu_e) }{1 + (\omega_e / \nu_e)^2} .
\label{eq:sigmah}
\end{equation}

In the limit of low ionization fraction, $\nu_{e}$ is closely related
to the mean collisional frequencies of the electrons with molecules of
the neutral gas such that (cf. Draine et al. 1983)
\begin{equation}
\nu_{e}^{(1)} = N_n <\sigma v>_{e-n}
\simeq N_n  \times 10^{-19}
       \left( \frac{128 k T}{9 \pi m_e} \right)^{1/2} .
\label{eq:nue1}
\end{equation}
where $N_n$ is the number density of neutral particles, $<\sigma
v>_{e-n}$ is the average product of collisional cross section and the
relative speed between electrons and neutral particles.  In the other
limit of a completely ionized gas, we take into account both
electron-ion and electron-electron encounters.  The collisional
frequency of the electrons is given by (cf. Sturrock 1994)
\begin{equation}
\nu_{e}^{(2)} \simeq 10^{8.0} N_e T^{-3/2}
\label{eq:nue2}
\end{equation}
In the intermediate range, we use $\nu_{e} = \max (\nu_{e}^{(1)}, 
\nu_{e}^{(2)})$. 

The resistance of the planet cannot be specified exactly because an
unknown shape factor is involved. Following Dermott (1970), we use the
following expression for the resistance:
\begin{equation}
R_p = \frac{f_R}{r_p^2} \int_0^{r_p} \frac{dr}{\sigma_p (r)} ,
\label{eq:resis}
\end{equation}
where $f_R$ is a order-of-unity parameter for the geometry.  In the
aligned geometry we consider here, the planet's resistance comes from
Pedersen resistivity (conductivity). The core of the planet is assumed
to be a perfect conductor. From Eq.~\ref{eq:resis}, we can see that it
is the cold, mostly neutral envelope that gives rise to most of the
resistance.

\end{document}